\documentclass[12pt,a4paper]{article}
\pdfoutput=1
\usepackage{a4wide}
\usepackage[centertags]{amsmath}
\usepackage{amssymb}
\usepackage{cite}
\usepackage{ulem}
\usepackage{float}
\usepackage[pdftex]{graphicx,color} 
\allowdisplaybreaks
\begin{document}

\thispagestyle{empty}

\begin{center}
{\Large  \bf Yukawa couplings for intersecting D-branes on
  non-factorisable tori} 
\end{center}

\vspace*{1cm}

\centerline{Stefan F{\"o}rste and Christoph Liyanage}

\vspace{1cm}

\begin{center}{\it
Bethe Center for Theoretical Physics\\
{\footnotesize and}\\
Physikalisches Institut der Universit\"at Bonn,\\
Nussallee 12, 53115 Bonn, Germany}
\end{center}

\vspace*{1cm}

\centerline{\bf Abstract}
\vskip .3cm
We compute Yukawa couplings in type IIa string theory compactified on
a six-torus in the presence of intersecting D6-branes. The six-torus
is generated by an SO(12) root lattice. Yukawa couplings are expressed
as sums over worldsheet instantons. Our result extends known expressions
to a non-factorisable torus. As an aside we also fill in some details
for the factorisable torus and non-coprime intersection numbers. 

\vskip .3cm

\newpage

\section{Introduction}

One of the avenues taken to find an embedding of Standard Model
particle physics into string theory consists of intersecting D6-branes
in type IIA theory 
\cite{Berkooz:1996km,Blumenhagen:1999ev,Forste:2000hx,Cvetic:2001tj,Cvetic:2001nr,Blumenhagen:2002gw,Cvetic:2002pj,Honecker:2003vq,Cvetic:2003xs,Blumenhagen:2003jy,Cvetic:2004ui,Honecker:2004kb,Honecker:2004np,Blumenhagen:2004xx,Blumenhagen:2005tn,Gmeiner:2005vz,Bailin:2006zf,Bailin:2007va,Bailin:2008xx,Gmeiner:2008xq,Forste:2010gw,Bailin:2011am,Honecker:2012qr,Honecker:2013hda,Ecker:2014hma}. For
reviews and a book providing also more references see
{e.g.} 
\cite{Angelantonj:2002ct,Blumenhagen:2005mu,Blumenhagen:2006ci,Ibanez:2012zz}.  
Toroidal orientifolds form a small subclass of more general
Calabi--Yau compactifications. Apart from their mathematical
simplicity the major advantage of toroidal models is that string
theory can be exactly quantised in those backgrounds. Various low
energy quantities can be computed from scratch. Effective degrees of
freedom are explicitly given in terms of string vibration
modes. Coupling constants are related to correlation functions of
corresponding vertex operators. Interactions in
intersecting brane models have been considered in
\cite{Aldazabal:2000cn,Cremades:2003qj,Cremades:2004wa,Abel:2002az,Cvetic:2003ch,Abel:2003fk,Abel:2003vv,Abel:2003yx,Lust:2004cx,Abel:2004ue,Abel:2005qn,Abel:2006yk,Duo:2007he,Russo:2007tc,Pesando:2012cx,Pesando:2014owa,Pesando:2014sca}. 

In the present paper we focus on worldsheet instanton contributions to
cubic Yukawa couplings. These have been worked out for a factorisable
six-torus in \cite{Cremades:2003qj}. In our context, a factorisable
six-torus is a six dimensional torus which can be decomposed into the
direct product of three mutually orthogonal two-tori. Yukawa couplings
can be written as product of three theta functions where each
function's arguments depend on the K{\"a}hler modulus of the
corresponding two-torus and open string moduli. The authors of
\cite{Cremades:2003qj} speculate that a more general
(non-factorisable)  six-torus will lead to multi-theta functions. We
will 
show this to be indeed the case by working out details for a
six-torus which is the quotient of ${\mathbb R}^6$ with respect to
shifts by SO(12) root vectors. Our analysis can be easily extended to
other non-factorisable six-tori. Intersecting D-branes on
non-factorisable tori and their orientifolds have been discussed in
\cite{Blumenhagen:2004di,Forste:2007zb,Kimura:2007ey,Forste:2008ex,
  Bailin:2013sya}.

In section 2 we recall general statements on how to obtain Yukawa
couplings for intersecting D6-branes. Section 3 discusses how to
represent cycles wrapped by D6-branes. Here we follow and supplement
the presentation of \cite{Forste:2007zb,Forste:2008ex}. A D6-brane
is, as in the factorisable case, described by three pairs of wrapping
numbers. The closeness to the factorisable representation turns out to
be of great advantage in the computation of Yukawa couplings. It is
also useful for constructing orientifolds where the point group leaves
also factorisable $T^6$ invariant. In other cases, such as the
${\mathbb Z}_{12}$ orientifold, a different representation of
D6-branes has been employed \cite{Blumenhagen:2004di,Bailin:2013sya}. It 
will be interesting to find out whether our way of characterising the 
wrapped three-cycle can be useful also for those constructions.  
Section 4 is devoted on how to label inequivalent intersection
points. It turns out that now such inequivalent labels take values on
a three dimensional lattice quotiented by a sublattice. 
In section 5 we derive a general expression for the Yukawa
coupling. It is expressed as a sum over a three dimensional
lattice resulting in a multi-theta function. In section 6 we illustrate
our general procedure at two examples. Section 7 provides our conclusions.
In an appendix we revisit a discussion carried out in
\cite{Cremades:2003qj}. We give a more detailed derivation of their
result for the case that intersection numbers are not coprime. 

\section{Intersecting D6-Branes and Worldsheet Instantons
\label{sec:2}}

Yukawa couplings within intersecting D6-brane models of type IIA
string theory have been discussed in general in
\cite{Cremades:2003qj}. There, for type IIA  strings compactified on a
factorisable six-torus, an explicit computation is presented. The
results can be easily extended to the, phenomenologically more
relevant, orientifolds of such tori by inclusion of respective image
branes. In our context, a factorisable six-torus is a torus which can
be decomposed into a product of three mutually orthogonal two-tori. We
are going to extend the results of \cite{Cremades:2003qj} to the case
of non-factorisable six-tori. Before doing so, let us recapitulate
some general considerations from \cite{Cremades:2003qj}.

After compactifing six of the
ten initial spacetime dimensions on a Calabi--Yau threefold
${\mathcal M}=CY^3$, the whole non-compact Minkowski space
$\mathbb{M}^4$ will be 
filled with D6-branes. In compact space they will wrap special
Lagrangian 3-cycles $\Pi_a \in H_3(\mathcal{M},\mathbb{Z})$: 
\begin{equation}
\mathbb{M}^4 \times \Pi_a \subset \mathbb{M}^4 \times \mathcal{M} .
\end{equation}
Compactification of type II theories then
leads to $\mathcal{N}=2$ supersymmetry in four dimensions. 
Breaking to the phenomenologically interesting amount of
$\mathcal{N}=1$ supersymmetry will be discussed shortly. 
Calabi--Yau threefolds are Ricci-flat K\"ahler manifolds
$\mathcal{M}$ with a complex structure $J$, a Riemannian metric
$g$ and additionally a nowhere vanishing holomorphic (3,0)-form
$\Omega$ called the volume form satisfying  
\begin{equation}\label{eq:om}
\frac{\omega^3}{3!} =-\left( \frac{i}{2}\right)^3 \Omega \wedge
\overline  \Omega ,
\end{equation}
where $\omega$ is the K\"ahler 2-form of $g$ and $J$. Both, the volume form
$\Omega$ and the K\"{a}hler 2-form $\omega$, are used to construct calibration
forms on $CY^3$. For the 3-cycles $\Pi_a$ to become special
Lagrangian, they must be calibrated by a real closed three form with
proper normalisation, {\it viz.} $\cos \theta\, \mathrm{Re} \Omega +
\sin\theta\, \mathrm{Im}\Omega =
\mathrm{Re}\left(\mathrm{e}^{-\mathrm{i}\theta}\Omega\right)$ 
(see e.g.\ \cite{Joyce-Book}). A calibrated 
submanifold is a respresentative of a given homology
class with minimal volume. For Calabi-Yau manifolds an equivalent
condition for submanifolds calibrated w.r.t.\  $\mathrm{Re}\left(
  \mathrm{e}^{-\mathrm{i}\theta}\Omega\right)$ is \cite{Harvey:1982xk} 
\begin{equation} \label{eq:calib}
\omega |_{\Pi_a} \equiv 0 \quad \text{and} \quad
\operatorname{Im}(e^{-i \theta}\Omega)|_{\Pi_a} \equiv 0 .
\end{equation}
The factor of $e^{-i \theta}$ is a phase, where $\theta$ parameterises how a
$U(1)_R$ symmetry of an $\mathcal{N}=1$ supersymmetry is embedded into
the $SU(2)_R$ of the, so far, unbroken $\mathcal{N}=2$
superalgebra. Indeed, placing a D-brane on a special Lagrangain cycle
breaks $\mathcal{N}=2$ to 
the corresponding $\mathcal{N}=1$ supersymmetry
\cite{Douglas:1999vm}. 

Due to Gauss' law the total RR charge generated by
D-branes (and O-planes in orientifolds) must add up to zero  in 
compact space. This is known as tadpole cancellation condition 
\begin{equation}
\sum_a N_a [\Pi_a]=0 .
\end{equation}
Stacking $N$ D-branes on top of each other open strings living on the
stack will form the adjoint representation of a $U(N)$ gauge
group. Each intersection of two stacks of D-branes with respective
gauge groups $U(N)$ and $U(M)$ accomodates chiral fermions in a
bifundamental representation of $U(N)\times U(M)$.  If the branes of
both stacks $\Pi_a$ and $\Pi_b$ 
are calibrated w.r.t.\ the same 3-form in (\ref{eq:calib}) (i.e.\
$\theta_a=\theta_b$) one scalar particle at each intersection point
becomes massless and fills together with the fermion a chiral ${\cal N}=1$
multiplet.  

In intersecting D-brane models one can have also topologically
non-trivial solutions to the e.o.m.\ of the open string, called a
worldsheet instanton. Open string instantons are Riemannian surfaces
embedded into the target space by placing the boundary of the surface
on 1-cycles lying in the compact factor of D-brane
worldvolumes. Yukawa couplings in Type IIA occur from open string
instantons 
connecting three intersection points where, for instance,  the Higgs
and a left- and 
right handed fermion live. To fulfill the e.o.m.\ the string instanton
must be a holomorphic disc $D$ calibrated w.r.t.\ $\omega$. The volume of
such an instanton can be associated with the classical value of the
string action and because this is minimised, the instanton volume is
also minimised. The area $A$ of $D$ in target space will determine
the size of the Yukawa couplings. For each triplet of intersection
points there can be infinitely many instanton solutions since the
Euclidean worldsheet can wrap two-cycles in compact space an
arbitrary integer number of times. Therefore, the Yukawa coupling,
$Y_{ijk}$, is expressed as an infinite sum over such wrapping numbers
\begin{equation}\label{eq:general}
Y_{ijk} \propto \sum e^{-\frac{A_{ijk}}{2\pi \alpha ^\prime}} ,
\end{equation}
where each component in $(i,j,k)$ labels inequivalent intersections of
a brane pair. So, the triplet stands for a triplet of chiral
multiplets. The individual labels differ among multiplets in the same
representation, {i.e.} they can be viewed as family indices. (Note,
however, that the number of families can in general depend on the
representation, unlike in the Standard Model).   
The area $A_{ijk}=\int_{\partial D}\omega$ is minimised, where $\partial D$ is the
the boundary's postion of the holomorphic disc in the target
space. The area depends on $(i,j,k)$ due to the 
condition that $\partial D$ has to pass through points which are
equivalent to the ones labeled by $(i,j,k)$.  

\section{Branes on Non-Factorisable Six-Tori}

Explicit computations by means of comformal field theory can be
performed when the Calabi-Yau 3-fold degenerates to an orbifold (on
which string theory is still well defined). For simplicity, we will
discuss the case of toroidal compactification. To obtain a
compactification we first decompose ten dimensional spacetime into a
product of four dimensional spacetime times ${\mathbb R}^6$. By
identifying points related by lattice shifts ${\mathbb R}^6$ is
replaced by a six-torus $T^6$. As an example for a non-factorisable
lattice we consider the $SO(12)$ root lattice, i.e.\ w.r.t.\ Cartesian
coordinates on ${\mathbb R}^6$  the compactification lattice is
generated by $SO(12)$ simple roots,
\begin{align}
e_1&= (1,-1,0,0,0,0) ,\nonumber \\
e_2&= (0,1,-1,0,0,0) ,\nonumber \\
e_3&= (0,0,1,-1,0,0) , \label{eq:latbas} \\
e_4&= (0,0,0,1,-1,0) , \nonumber \\
e_5&= (0,0,0,0,1,-1) , \nonumber \\
e_6&= (0,0,0,0,1,1) .\nonumber
\end{align}
It is convenient to introduce three complex coordinates on ${\mathbb
  R}^6$, as
\begin{equation}\label{eq:cc}
z_i=x_{2i-1}+\text{i} x_{2i} ,\,\,\, i=1,2,3.
\end{equation}
So, the K\"{a}hler two-form is
\begin{equation}
\omega=\frac{i}{2}\sum_{i=1}^3 \operatorname{d}z_i \wedge
\operatorname{d} \overline{z}_i .
\end{equation}
The associated $(3,0)$-form satisfying (\ref{eq:om}) is 
\begin{equation}
\Omega = dz_1 \wedge dz_2 \wedge dz_3 .
\end{equation}
Note, that (\ref{eq:om}) fixes $\Omega$ only up to a phase. This
ambiguity is already parameterised by $\theta$ in the calibration form
$\mathrm{Re}\left(
  \mathrm{e}^{-\mathrm{i}\theta}\Omega\right)$. Another ambiguity
lies in the choice of complex structure, i.e.\ by picking the pairs of
real coordinates which form a complex one in
(\ref{eq:cc}). Here, we anticipate that finally one wants to
take an orbifold of $T^6$ where the point group consists of
simultaneous rotations in the three complex planes such that the
rotation angles add up to zero. Our choice ensures that $\omega$ as
well as $\Omega$ are invariant under the point group. 
Now, we need to specify special Lagrangian submanifolds w.r.t.\
$\mathrm{Re}\left( 
  \mathrm{e}^{-\mathrm{i}\theta}\Omega\right)$. Viewed as submanifolds
of ${\mathbb C}^3$, they should have the additional property that
constant shifts in the coordinates 
provide again special Lagrangian submanifolds. This ensures that we
can compactify ${\mathbb C}^3$ to a $T^6$. 
The first condition in (\ref{eq:calib}) is satisfied if the location of
the D-brane can be obtained from three equations relating $z_i$ to its
complex conjugate for each $i=1,2,3$. In particular we take the
submanifold of solutions to
\begin{equation}\label{eq:lines}
x_{2i} =\left(\tan \varphi_i\right)\, x_{2i-1},\,\,\, i=1,2,3.
\end{equation}
Eq.\ (\ref{eq:lines}) defines angles up to shifts by $\pi$. However, we define
the angles up to shifts by $2\pi$ by encoding also the orientation of
a line via the sign of $\cos\varphi_i$ and, if $\cos\varphi_i =0$, via the
sign of $\sin\varphi_i$.
The second condition in (\ref{eq:calib}) leads to
\begin{equation}\label{eq:imcon}
\sin\left( \theta -\varphi_1 -\varphi_2 -\varphi_3\right) = 0,
\end{equation}
and thus specifies the form w.r.t.\ which our submanifold is
calibrated. Indeed the calibration form on our submanifold is
\begin{equation}\label{eq:cafo}
\left.\mathrm{Re}\left(
    \mathrm{e}^{-\mathrm{i}\theta}\Omega\right)\right|_{\Pi} =
\frac{\cos\left( \theta -\varphi_1 -\varphi_2
    -\varphi_3\right)}{\cos\varphi_1\, \cos\varphi_2\, \cos\varphi_3}
dx_1 \wedge dx_2 \wedge dx_3 .
\end{equation}
If $\left|\varphi_i\right| = \pi/2$ we replace $dx_i$  by $\left(\cot
  \varphi_i\right)\, 
dy_i$. 
The metric induced on the submanifold is
\begin{equation}
\left. g\right|_\Pi = \sum_{i=1}^3 \left( 1 +\tan^2 \varphi_i\right)
dx_i ^2
\end{equation}
leading to the volume form
\begin{equation}\label{eq:volfo}
\left.\mathrm{vol}\right|_\Pi = \sqrt{\det g}\, dx_1 \wedge dx_2 \wedge
dx_3 =\frac{1}{\cos \varphi_1\, \cos\varphi_2
    \cos\varphi_3 }dx_1\wedge dx_2 \wedge dx_3 ,
\end{equation}
where, in the second step, we have chosen the branch of the square root
such that shifting individual angles by $\pi$ reverses the
orientation, i.e.\ changes the sign of the volume form. 
Equating (\ref{eq:volfo}) to the expression in (\ref{eq:cafo}) removes
one of the 
two solutions to (\ref{eq:imcon}), and yields
\begin{equation}
\theta = \varphi_1 +  \varphi_2  + \varphi_3 .
\end{equation}
Shifting $\theta$ by $\pi$ produces and extra sign in the calibration
form which can be absorbed by inverting the orientation on the
submanifold.  A D-brane on a submanifold of one orientation is
equivalent to an anti D-brane on the same submanifold with inverted
orientation. 

In summary, we can 
parametrise the 3-cycle homology class as in the factorisable case
\begin{eqnarray}
\Pi_a=\prod_{i=1}^3(n_a^i[a] + m_a^i[b]) &\text{with}& n_a^i,
m_a^i \in \mathbb{Z}  \label{eq:cyl}
\end{eqnarray}
where
\begin{equation}\label{eq:bascyc}
[a] =(1,0)\,\,\, ,\,\,\, [b]= (0,1)
\end{equation}
are one-cycles on a $T^2 = {\mathbb R}^2/\Lambda$, and $\Lambda$ is a
quadratic lattice of unit size.
%
The subscript $a$ labels different cycles (to be wrapped by
branes). The previously used angles can be obtained from
\begin{equation}
\varphi_i = \arctan \frac{m^i}{n^i} ,\,\,\,\begin{array}{lll}
\mathrm{sign}\left( \cos \varphi_i\right) = \mathrm{sign}\left(
  n^i\right)& \mbox{if} &  n^i \not= 0,\\ \mathrm{sign}\left(
  \sin \varphi_i\right) = \mathrm{sign}\left( 
 m^i\right)&\mbox{if} & n^i = 0 ,
\end{array}
\end{equation}
where we supressed the label $a$. Later, we will
turn on geometric moduli in each plane (such that the notion of
factorisable versus non-factorisable is preserved). This is done by
modifying (\ref{eq:bascyc}) to ($i\in \left\{ 1,2,3\right\}$)
\begin{equation} \label{eq:defcyc}
[a]^i = \left( a_1 ^i ,a_2^i\right) \,\,\, ,\,\,\, [b]^i = \left( b_1
  ^i,b_2 ^i\right) .
\end{equation}
The value of the angles is now computed from
\begin{equation}
\varphi_i = \arctan \frac{ n^i a_2^i + m^ib_2^i}{n^i a_1 ^i + m b_1 ^i} ,
\,\,\,\begin{array}{lll}
\mathrm{sign}\left( \cos \varphi_i\right) = \mathrm{sign}\left(
  n^ia_1^i + m^i b_1^i\right)& \mbox{if} &  n^ia_1^i + m^i b_1^i \not=
0,\\ \mathrm{sign}\left( 
  \sin \varphi_i\right) = \mathrm{sign}\left( 
 n^ia_2^i + m^i b_2^i\right)&\mbox{if} & n^ia_1^i + m^i b_1^i = 0 .
\end{array}
\end{equation}
At first, (\ref{eq:cyl}) looks applicable only to factorisable
six-tori. The product is to be understood as the cross       
product\footnote{The crossproduct of three subsets $A$, $B$ and $C$
  of a set $M$ is a 
  subset of the Cartesian product $M\times M\times M$ given by
  $A\times B\times C = \left\{ (a,b,c)\left| a\in A, b\in B, c\in
      C\right.\right\}$.} mapping 
a triplet of one-cycles in $T^2$ to a three-cycle in $T^2\times T^2
\times T^2$.  However, with some modifications this notion can also be
carried over to the non-factorisable $T^6 = {\mathbb
  R}^6/\Lambda_{\mathrm{SO(12)}}$ \cite{Forste:2007zb}\footnote{In
  \cite{Forste:2007zb} a three-cycle was associated with the
  antisymmetrised tensor product of SO(12) lattice vectors. Here, we
  just 
  focus on the set of points representing the cycle.}. Most easily
this can be seen if (we drop the brane label $a$ in the present discussion)
\begin{equation}\label{eq:close}
m^i + n^i = 0 \,\,\,\text{mod}\,\,\, 2
\end{equation}
for all $i=1,2,3$. In this case the cross product yields trivially a
closed three-cycle on  ${\mathbb
  R}^6/\Lambda_{\mathrm{SO(12)}}$. Consider now the case that for one
plane, say the first, the sum of the wrapping numbers is odd, i.e.\
\begin{equation}
m^1 + n^1 = \text{odd} .
\end{equation}
We claim that then the cross product
\begin{equation}\label{eq:cross}
(2n^1,2m^1)\times (n^2,m^2) \times (n^3,m^3)
\end{equation}
yields a closed three-cycle on ${\mathbb
  R}^6/\Lambda_{\mathrm{SO(12)}}$ for any integer choice of the
wrapping numbers in the second and third plane. The cross product
(\ref{eq:cross}) can be viewed as a set of points
\begin{equation}\label{eq:cycset}
M =\left\{ \left( 2n^1 x, 2 m^1 x, n^2 y, m^2 y, n^3 z , m^3 z\right)
  ,\,\,\, x,y,z \in \left[ 0,1\right]\right\} .
\end{equation}
It remains to show that this set is an invariant subset of ${\mathbb
  R}^6/\Lambda_{\mathrm{SO(12)}}$ under shifting any of
the parameters $x,y,z$ by one. For $x$ this is trivial. If $n^2 + m^2$
is odd the statement can be seen also for $y \to y+1$ after a suitable
reparameterisation of our set $M$. To this end we first decompose the
set
\begin{align}
M &= M_1 \cup M_2 , \nonumber\\
M_1 &= \left\{ \left( n^1 x,  m^1 x, n^2 y, m^2 y, n^3 z , m^3 z\right)
  ,\,\,\, x,y,z \in \left[ 0,1\right]\right\} ,\\
M_2 &=  \left\{ \left( n^1 x + n^1,  m^1 x + m^1, n^2 y, m^2 y, n^3 z , m^3 z\right)
  ,\,\,\, x,y,z \in \left[ 0,1\right]\right\} . \nonumber
\end{align}
Since both, $n^1 + m^1$ and $n^2 + m^2$ are odd the six dimensional
vector
\begin{equation}
\vec{V} =\left( -n^1 , -m^1, n^2 ,m^2 ,0,0\right)
\end{equation}
is in the SO(12)  lattice. Shifting the position vector in
$M_2$ by that vector and forming the union of $M_1$  with
$M_2$ afterwards we obtain an equivalent parameterisation of $M$ in
${\mathbb R}^6/\Lambda_{\mathrm{SO(12)}}$,
\begin{equation}
M = \left\{ \left( n^1 x,  m^1 x, 2 n^2 y, 2 m^2 y, n^3 z , m^3 z\right)
  ,\,\,\, x,y,z \in \left[ 0,1\right]\right\},
\end{equation}
in which closure under $y \to y+ 1$ is obvious. If $m^3 + n^3$ is also
odd an analogous argument shows closure under $z \to z+1$. 
In summary, we characterise a three-cycle on ${\mathbb
  R}^6/\Lambda_{\mathrm{SO(12)}}$ by three pairs of integer wrapping
numbers $(n^i,m^i)$, $i=1,2,3$ such that one of the following
conditions hold:
\begin{itemize} \label{list:cases}
\item Condition (\ref{eq:close}) holds for $i=1,2,3$. For at least two
  pairs $n^i$ and $m^i$ are coprime. For the remaining pair they are
  either also coprime or can be written as $n^i = 2k$, $m^i = 2l$
  with $k+l$ odd and $k$, $l$ coprime.
\item Condition (\ref{eq:close}) does not hold for exactly one
  pair. The integers in that pair are coprime. One of the pairs
  satisfying (\ref{eq:close}) contains coprime integers whereas the
  other one consists of even numbers $\left(2k,2l\right)$ with coprime
  integers $k,l$ and $k+l$ is odd.
\item Condition (\ref{eq:close}) does not hold for exactly two
  pairs. The remaining pair consists of even numbers
  $\left(2k,2l\right)$ with coprime integers $k,l$ and $k+l$ is odd.
\end{itemize}
Otherwise the cycle is either multiply wrapped or not closed in
${\mathbb R}^6/\Lambda_{\mathrm{SO(12)}}$. For instance the cycle $(2,0)
\times (2,0)\times (1,0)$ corresponds to a stack of two branes
wrapping $(2,0)\times (1,0) \times (1,0)$, for which the last
condition of our list is satisfied. On the other hand, the cycle
$(1,0) \times (1,0) \times (1,0)$ is not closed and has to be replaced
by e.g.\ $(2,0)\times (1,0)\times (1,0)$.   

\section{Labelling Inequivalent Intersections}

In order to write down general expressions (such as in
({\ref{eq:general})) it is useful to label different chiral multiplets
  by different numbers. Since chiral multiplets are localised at
  intersection points of two stacks of D-branes this amounts to
  labelling inequivalent intersection points. Let us recall how this is
  done for a factorisable $T^6$. Consider two stacks of branes
  wrapping three-cycles $\Pi_a$ and $\Pi_b$, respectively. These
  three-cycles are specified by their wrapping numbers as in
  (\ref{eq:cyl}), respectively an analogous expression with subscript
  $a$ replaced by $b$. For the factorisable $T^6$ the intersection
  number can be written as a product of three integers ({\it viz.} the
    intersection numbers on the $T^2$ factors)
\begin{equation}\label{eq:trip}
I_{ab} = \prod_{i=1}^3 I^{(i)}_{ab},\,\,\,\mbox{with\ } I^{(i)}_{ab} =
n_a^i m_b^i - m_a^i n_b^i .
\end{equation}
Therefore, it is convenient to label the intersecction point by a
triplet of integers
\begin{equation}
\left( j^{(1)}, j^{(2)} , j^{(3)}\right) , \,\,\, \mbox{with\ } j^{(i)} \in \left\{
  0,1,\ldots , \left|I_{ab}^{(i)}\right| -1\right\} .
\end{equation}
It remains to associate a particular intersection locus to a
label. For the factorisable $T^6$ we focus on a $T^2$  factor
which we obtain by taking the quotient of the complex plane with
respect to a square unit lattice. 
For the moment, we consider the case that all D-branes pass through
the origin. The equation determining the intersection points is
\begin{equation} \label{eq:faceq}
\left( \begin{array}{c} n_a\, x \\ m_a\, x \end{array}\right) =
\left( \begin{array}{c} n_b\, y \\ m_b\, y\end{array}\right) +
\left( \begin{array}{c} k \\ l \end{array}\right) ,
\end{equation}
where $k$ and $l$ are integers corresponding to lattice shifts. The
real variables $x$ and $y$ parameterise the subsets at which the
D-branes are located. One finds the following positions of
intersection points
\begin{equation}
{\Pi_a \cap \Pi_b}_{\left| T^2\right.} =\left\{
  \left.\left( \begin{array}{c} n_a \frac{km_b - 
        ln_b}{I_{ab}} \\ 
m_a \frac{k m_b - l n_b}{I_{ab}}\end{array}\right)\right|\,\,\, k,l \in
{\mathbb Z}\right\} .
\end{equation}
If the factor $\frac{k m_b - l n_b}{I_{ab}}$ is integer the
corresponding intersection point is equivalent to the origin. Since
$m_a$ and $n_a$ are coprime $I_{ab}$ cannot be a non
trivial divisor of both numbers. The integers $k$ and $l$ specify the
position of the intersection point. The relation to its label is given
by
\begin{equation}\label{eq:labfac}
k m_b - l n_b = i\,\,\, \mbox{mod}\,\,\, \left|I_{ab}\right| .
\end{equation}
Notice that since $n_b$ and $m_b$ are coprime $k$ and $l$ can be
arranged such that the left hand side of (\ref{eq:labfac}) equals any
of the integers in $\left[ 0,\left| I_{ab}\right| \right)$. 
For each of the $T^2$  factors the space of inequivalent labels is
given by 
$$ \frac{{\mathbb Z}}{\left| I_{ab}\right|{\mathbb Z}} . $$
Combining the three $T^2$  factors making up the factorisable $T^6$ we
find that labels of inequivalent intersection points are triplets
taking values on the factorisable three dimensional quotient lattice
\begin{equation}
\left( i^{(1)}, i^{(2)}, i^{(3)}\right) \in \bigotimes_{k=1}^3
\frac{{\mathbb Z}}{\left| I^{(k)}_{ab}\right|{\mathbb Z}} .
\end{equation}

For the non-factorisable $T^6$ we will see now that inequivalent
labels can take values on a non-factorisable three dimensional quotient lattice.
Instead of (\ref{eq:faceq}) we have 
to solve the full six dimensional equation and find (up to SO(12)
lattice shifts) 
\begin{align}
\Pi_a \cap \Pi_b =\Bigl\{ &\left( A_{ab}\, n_a^1 , A_{ab}\,
  m_a^1,  B_{ab}\, n_a^2 , 
    B_{ab}\, m_a^2,  C_{ab}\, n_a^3 , C_{ab}\, m_a^3\right)^{\mathrm{T}},  \,\,\,
  \mbox{with}  \nonumber\\
 &A_{ab} =  \frac{t_1m_b^1-t_2n_b^1}{I_{ab}^{(1)}} ,\,\, B_{ab} =
 \frac{t_3m_b^2-t_4n_b^2}{I_{ab}^{(2)}} ,\,\, C_{ab} =
 \frac{t_5m_b^3-t_6n_b^3}{I_{ab}^{(3)}}, \label{eq:intpo}\\ 
 & \left. I_{ab}^{(i)}=\left(n_a^im_b^i-n_b^im_a^i\right),\,\,\,
 \vec{t}=\left(t_1,t_2,t_3,t_4,t_5,t_6\right)^{\mathrm
{T}}\in
 \Lambda_{\mathrm{SO(12)}} \right\}\nonumber 
\end{align}
The actual space of inequivalent labels depends on which of the
configurations listed in section \ref{list:cases} is
realised. We will discuss nine different examples in the following.

\subsection*{Case 1}

Here, we consider the case that 
\begin{equation}
\forall_{i \in \left\{ 1,2,3\right\}} \forall_{x \in \left\{a,b\right\}} \left( n_x
  ^i + m_x ^i =  0\, \text{mod}\, 2,\,\,\, g.c.d.\left( n_x^i, m_x
    ^i\right) =1\right) ,
\end{equation}
where $g.c.d.\left( \ldots\right)$  assigns the greatest common divisor
to the list of integers in its argument. 
The inequivalent labels are in a subset of
\begin{equation}\label{eq:indlat}
\left\{ \left( t_1m_b^1-t_2n_b^1 , t_3m_b^2-t_4n_b^2 ,
    t_5m_b^3-t_6n_b^3\right) |  \vec{t} \in
  \Lambda_{\mathrm{SO(12)}}\right\} = \Lambda_{\text{SO(6)}},
\end{equation}
where $\Lambda_{\text{SO(6)}}$  is the lattice generated by SO(6)
simple roots, 
\begin{equation}
\alpha_1 = \left( 1, -1, 0\right) ,\, \alpha_2 =\left( 0,1,-1\right)
,\, \alpha_3  = \left( 0,1,1\right) .
\end{equation}
Shifting the $i^{\text{th}}$ component of the three-dimensional label
by an integer multiple of $I^{(i)}_{ab}$  leads to an equivalent
label. Thus our quotient lattice of inequivalent labels is
\begin{equation}
\frac{\Lambda_{\text{SO(6)}}}{I_{ab}^{(1)} {\mathbb Z} \otimes
 I_{ab}^{(2)} {\mathbb Z} \otimes  I_{ab}^{(3)} {\mathbb Z}}
\end{equation}
The number of inequivalent intersection points is given by the index of the
quotient lattice
\begin{equation}
\left| \frac{\Lambda_{\text{SO(6)}}}{I_{ab}^{(1)} {\mathbb Z}\otimes
 I_{ab}^{(2)} {\mathbb Z} \otimes I_{ab}^{(3)}  {\mathbb Z}}
\right| =  \left| \frac{E_{I_{ab}^{(1)} {\mathbb Z} \otimes
I_{ab}^{(2)}  {\mathbb Z} \otimes I_{ab}^{(3)}  {\mathbb
    Z}}}{E_{\Lambda_{\text{SO(6)}}}}\right| ,
\end{equation}
where the $E$'s are the determinants of the dreibein corresponding to
the lattice indicated by the subscript. Especially  
\begin{align}
E_{I_{ab}^{(1)} {\mathbb Z} \otimes
I_{ab}^{(2)}  {\mathbb Z} \otimes I_{ab}^{(3)} {\mathbb Z}}& =
\det \left( \begin{array}{ccc} 
I_{ab}^{(1)} & 0 & 0\\
0 & I_{ab}^{(2)} & 0\\
0 & 0 & I_{ab}^{(3)} 
\end{array}
\right) =  I_{ab}^{(1)} I_{ab}^{(2)} I_{ab}^{(3)} , \\
E_{\Lambda_{\text{SO(6)}}} & = \det \left( \begin{array}{ccc} 1 & -1 &
    0 \\ 0 & 1 & -1\\ 0 & 1 & 1 \end{array}\right) = 2 .
\end{align}
Hence the number of inequivalent intersection labels equals the
modulus of the intersection number $I_{ab}$, where
\cite{Blumenhagen:2004di,Forste:2007zb} 
\begin{equation} \label{eq:inum}
I_{ab} = \frac{1}{2}\prod_{i=1}^3 I_{ab}^{(i)} .
\end{equation}
As an aside we notice that in the case considered here all the
$I_{ab}^{(i)}$  are even which can be easily seen by rewriting
\begin{equation}\label{eq:even}
I_{ab}^{(i)} = \frac{  n_a  ^i - m_a ^i}{2}\left( n_b ^i + m_b
  ^i\right) - \frac{n_a ^i +m_a ^i}{2}\left( n_b ^i - m_b ^i\right)
, 
\end{equation}
which is the difference between two even numbers. This confirms that
we are taking the quotient with respect to a sublattice of
$\Lambda_{\text{SO(6)}}$. 

\subsection*{Case 2}

Here we consider the configuration
\begin{align}
& g.c.d.\left( n_a^1 , m_a^1\right)  = 2 ,\,\,\, \frac{n_a ^1 + m_a ^1}{2}
  = \text{odd}, \nonumber\\
& \forall_{i \in \left\{2,3\right\}} \left( n_a
  ^i + m_a ^i =  0\, \text{mod}\, 2,\,\,\, g.c.d.\left( n_a^i, m_a
    ^i\right) =1\right) ,  \\
& \forall_{i \in \left\{ 1,2,3\right\}} \left( n_b
  ^i + m_b ^i =  0\, \text{mod}\, 2,\,\,\, g.c.d.\left( n_b^i, m_b
    ^i\right) =1\right) .  \nonumber
\end{align}
Since the configuration of stack $b$ is the same as in the previous
case conclusion (\ref{eq:indlat}) still holds. 
Now shifting $i^{(1)}$ by an integer multiple of $I_{ab}^{(1)}/2$
already leads to an equivalent point on $T^6$. One can show that
$I_{ab}^{(1)}/2$ can be written as the difference of an even and an
odd number and hence is odd. So 
shifting $i^{(1)}$  by $I_{ab}^{(1)}/2 $ has to be accompanied by
shifting any of the other two remaining label components by an odd
number. This however leads to an inequivalent point since
$I_{ab}^{(2)}$ and $I_{ab}^{(3)}$ are even. Therefore the sublattice
with respect to which we take the quotient is the same as in the
previous example and we arrive at the same result.  

\subsection*{Case 3}

Here, we  consider the previous situation with the roles of stacks $a$
and $b$ swapped,
\begin{align}
& \forall_{i \in \left\{ 1,2,3\right\}} \left( n_a
  ^i + m_a ^i =  0\, \text{mod}\, 2,\,\,\, g.c.d.\left( n_a^i, m_a
    ^i\right) =1\right) ,  \nonumber\\
& g.c.d.\left( n_b^1 , m_b^1\right)  = 2 , \,\,\, \frac{n_b ^1 + m_b ^1}{2}
  = \text{odd},\\
& \forall_{i \in \left\{2,3\right\}} \left( n_b
  ^i + m_b ^i =  0\, \text{mod}\, 2,\,\,\, g.c.d.\left( n_b^i, m_b
    ^i\right) =1\right) .  \nonumber
\end{align}
Now, the label takes values in 
\begin{equation}
\left\{ \left( t_1m_b^1-t_2n_b^1 , t_3m_b^2-t_4n_b^2 ,
    t_5m_b^3-t_6n_b^3\right) |  \vec{t} \in
  \Lambda_{\mathrm{SO(12)}}\right\}= 2 {\mathbb Z}\otimes {\mathbb
  Z}\otimes {\mathbb Z} 
\end{equation}
where $2{\mathbb Z}$  denotes the set of even numbers. The difference
to (\ref{eq:indlat}) comes about as follows. Since $n_b^1$ and $m_b^1$
are even the first component of the index is even. It   is not
affected by shifts
\begin{equation} \label{eq:inco}
\left( \begin{array}{c} t_1 \\ t_2\end{array}\right) \to
\left( \begin{array}{c} t_1 \\ t_2\end{array}\right) +
\left(  \begin{array}{c}
\frac{n_b ^1}{2}\\ \frac{m_b ^1}{2}\end{array}\right) .
\end{equation}
This shift does not correspond to the first two components of a
$\Lambda_{\text{SO(12)}}$  lattice vector. So, if we shift any of the
other label-components by an integer we can always associate a
$\Lambda_{\text{SO(12)}}$ lattice vector to it  by leaving or adding the
  contribution (\ref{eq:inco}). So, now we obtain for the space of
  inequivalent labels the quotient lattice
\begin{equation}\label{eq:lat3}
\frac{ 2{\mathbb Z}\otimes {\mathbb Z}\otimes  {\mathbb Z}}{I_{ab}^{(1)} 
{\mathbb Z} \otimes 
 I_{ab}^{(2)} {\mathbb Z} \otimes  I_{ab}^{(3)} {\mathbb Z}} ,
\end{equation}
whose index is again $\left|I_{ab}\right|$  as it should be.   

\subsection*{Case 4}

The fourth configuration we consider is
\begin{align}
& g.c.d.\left( n_a^1 , m_a^1\right)  = 2 ,\,\,\, \frac{n_a ^1 + m_a ^1}{2}
  = \text{odd}, \nonumber\\
& \forall_{i \in \left\{2,3\right\}} \left( g.c.d.\left( n_a^i, m_a
    ^i\right) =1\right) ,\,\,\, n_a ^2 + m_a ^2 = \text{even},\,\,\,
n_a ^3 + m_a ^3 =\text{odd},  \\
& \forall_{i \in \left\{ 1,2,3\right\}} \left( n_b
  ^i + m_b ^i =  0\, \text{mod}\, 2,\,\,\, g.c.d.\left( n_b^i, m_b
    ^i\right) =1\right) .  \nonumber
\end{align}
Since the configuration of stack $b$ is the same as in the first case
the space of inequivalent labels will be again some quotient of
$\Lambda_{\text{SO(6)}}$. This time  the sublattice of equivalence
shifts looks slightly different, it is generated by 
\begin{equation} \label{eq:betas}
\beta_1 =\left( I_{ab}^{(1)}, 0,0\right),\,\,\, \beta_2 =\left( 0,
  I_{ab}^{(2)},0\right) ,\,\,\, \beta_3 =\left( \frac{1}{2}
  I_{ab}^{(1)},0, I_{ab}^{(3)}\right) .
\end{equation}
The determinant of this dreibein is the same as in case one and hence
the index of the quotient lattice is still $\left| I_{ab}\right|$.

\subsection*{Case 5}

We consider the previous case with $a$ and $b$  interchanged,
\begin{align}
& \forall_{i \in \left\{ 1,2,3\right\}} \left( n_a
  ^i + m_a ^i =  0\, \text{mod}\, 2,\,\,\, g.c.d.\left( n_a^i, m_a
    ^i\right) =1\right) ,  \nonumber\\
& g.c.d.\left( n_b^1 , m_b^1\right)  = 2 ,\,\,\, \frac{n_b ^1 + m_b ^1}{2}
  = \text{odd}, \\
& \forall_{i \in \left\{2,3\right\}} \left( g.c.d.\left( n_b^i, m_b
    ^i\right) =1\right) ,\,\,\, n_b ^2 + m_b ^2 = \text{even},\,\,\,
n_b ^3 + m_b ^3 =\text{odd}.  \nonumber
\end{align}
The quotient lattice of inequivalent intersection labels is the same
as in (\ref{eq:lat3}).

\subsection*{Case 6}

The set of wrapping numbers satisfies
\begin{align}
& \forall_{x\in \left\{ a,b\right\}} \left( g.c.d.\left( n_x^1,
    m_x^1\right) = 2,\,\,\, \frac{ n_x^1 + m_x^1}{2}
  = \text{odd}\right) , \nonumber \\
&\forall_{x\in \left\{ a,b\right\}} \forall_{i\in \left\{
    2,3\right\}}\left( n_x ^i + m_x ^i = 0 \,\text{mod}\,
2,\,\,\, g.c.d.\left( n_x^i, m_x^i\right) =1\right) .
\end{align}
Inequivalent labels are in (\ref{eq:lat3}). This does not change if we
permute the three pairs of wrapping numbers for stack $a$ whereas
such a permutation in stack $b$ results in a corresponding permuation
of the order of the factors in the numerator in (\ref{eq:lat3}).

\subsection*{Case 7}

We consider configurations of the form
\begin{align}\
& g.c.d.\left( n_a ^1, m_a ^1\right) =2,\,\,\, \frac{n_a^1 + m_a^1}{2}
=\text{odd} , \nonumber \\
& \forall_{i\in \left\{2,3\right\}} \left( n_a^i + m_a ^i =
  0\,\text{mod}\, 2,\,\,\, g.c.d.\left( n_a^i,m_a^I\right) =1\right)
,\nonumber \\
& g.c.d.\left( n_b ^1, m_b ^1\right) =2,\,\,\, \frac{n_b^1 + m_b^1}{2}
=\text{odd} , \\ 
& \forall_{i\in \left\{2,3\right\}} \left(g.c.d.\left(
    n_b^i,m_b^i\right) =1\right) , \nonumber \\
& n_b ^2 + m_b ^2 = 0 \,\text{mod}\, 2,\,\,\, n_b^3 + m_b^3 =
\text{odd} .
\nonumber
\end{align}
The space of inequivalent labels is again quotient lattice
(\ref{eq:lat3}). Permuting the pairs of wrapping number in stack $a$ does
not change this. Permuting the pairs of wrapping numbers in stack $b$
alters the position of $2{\mathbb Z}$ in (\ref{eq:lat3}). 

\subsection*{Case 8}

We take the configuration of the previous case with $a$ and $b$
swapped,
\begin{align}
& g.c.d.\left( n_a ^1, m_a ^1\right) =2,\,\,\, \frac{n_a^1 + m_a^1}{2}
=\text{odd} ,\nonumber \\ 
& \forall_{i\in \left\{2,3\right\}} \left(g.c.d.\left(
    n_a^i,m_a^i\right) =1\right) , \nonumber \\
& n_a ^2 + m_a ^2 = 0 \,\text{mod}\, 2,\,\,\, n_a^3 + m_a^3 =
\text{odd} ,
\\
& g.c.d.\left( n_b ^1, m_b ^1\right) =2,\,\,\, \frac{n_b^1 + m_b^1}{2}
=\text{odd} , \nonumber \\
& \forall_{i\in \left\{2,3\right\}} \left( n_b^i + m_b ^i =
  0\,\text{mod}\, 2,\,\,\, g.c.d.\left( n_b^i,m_b^i\right) =1\right)
.\nonumber
\end{align}
The space of inequivalent labels are equivalence classes in the set of
lattice vectors 
\begin{equation}
2{\mathbb Z}\otimes {\mathbb Z}\otimes {\mathbb Z} ,
\end{equation}
with respect to shifts generated by
\begin{equation}
\beta_1 =\left( I_{ab}^{(1)}, 0,0\right) ,\,\,\, \beta_2=\left(
  \frac{I_{ab}^{(1)}}{2}, I_{ab}^{(2)}, 0\right) ,\,\,\, \beta_3
=\left( 0,0, I_{ab}^{(3)}\right).
\end{equation}
The index of this quotient lattice is $\left| I_{ab}\right|$. 
Permuting pairs of wrapping numbers gives again very similar results
and will not be discussed in detail.

\subsection*{Case 9}

Finally, we look at configurations of the form
\begin{align}   
& g.c.d.\left( n_a ^1, m_a ^1\right) =2,\,\,\, \frac{n_a^1 + m_a^1}{2}
=\text{odd} ,\nonumber \\ 
& \forall_{i\in \left\{2,3\right\}} \left(g.c.d.\left(
    n_a^i,m_a^i\right) =1\right) , \nonumber \\
& n_a ^2 + m_a ^2 = 0 \,\text{mod}\, 2,\,\,\, n_a^3 + m_a^3 =
\text{odd} ,
\\
& g.c.d.\left( n_b ^1, m_b ^1\right) =2,\,\,\, \frac{n_b^1 + m_b^1}{2}
=\text{odd} ,\nonumber \\ 
& \forall_{i\in \left\{2,3\right\}} \left(g.c.d.\left(
    n_b^i,m_b^i\right) =1\right) , \nonumber \\
& n_b ^2 + m_b ^2 = 0 \,\text{mod}\, 2,\,\,\, n_b^3 + m_b^3 =
\text{odd} .
\nonumber
\end{align}
Here, the space of inequivalent labels is identical to the previous
case 8.

\section{Yukawa Couplings \label{sec:yuk}}

In \cite{Cremades:2003qj} Yukawa couplings have been computed as a sum
over worldsheet instantons. These instantons are Euclidean open string
worldsheets. The worldsheet parameters take values on a disc which is
holomorphically or anti-holomorphically mapped to a two dimensional
surface in target space. The boundary of the disc is mapped to curves
lying within 
the wordlvolume of D6-branes and passing through three intersection
points. The three 
multiplets, whose Yukawa coupling is being computed, are each
localised at one of the three intersection points. As  explained in
\cite{Cremades:2003qj} (especially in their appendix A\footnote{The
  arguments given there do not depend on the choice of
  compactification lattice.}) projections of the worldsheet image onto
each of the three complex planes (appearing in $T^6 = {\mathbb
  C}^3/\Lambda_{\mathrm{SO(12)}}$) are straight triangles (possibly
points). The instanton action is the area of the worldsheet
image in units of $\alpha^\prime$. For a holomorphic or
anti-holomorphic embedding this area is given by the sum of the areas
of the three projection triangles (where a point has zero area). 

Here, we will also include the possibility of brane stacks not passing
through the origin, i.e.\ we shift the branes position by
\begin{equation}
\Pi_x \to \Pi_x + \sum_{i=1}^3 \frac{\epsilon_x
  ^i}{\left(n_x^i\right)^2 +\left( m_x ^i\right)^2} \left( -m_x ^i
  \mathbf{k}_{2i-1} + n_x 
  ^i \mathbf{k}_{2i}\right) ,
\label{eq:epsdef}
\end{equation}
where the $\mathbf{k}_l$'s form the canonical basis of ${\mathbb R}^6$
(one in the $l^{\text{th}}$ row and zero else) and $x\in \left\{
    a,b,c\right\}$. The positions of intersection points are shifted
  by the $\epsilon$'s. For the loci of the triangle's vertices in the 
  $i^\text{th}$ plane we find\footnote{Although many details of the
    calculation do not differ from the factorisable case 	 	
    \cite{Cremades:2003qj} we present it for the sake of our presentation's
    self-containedness.}
\begin{eqnarray}
\left( ab\right)_i &=& \left(\begin{array}{c} n^i_b \\
    m^i_b\end{array}\right) \left(\frac{i^{(i)}}{I_{ba}^{(i)}} +\frac{
    \epsilon^i_a}{I^{(i)}_{ab}} -\frac{\left( m^i_b m^i_a + n^i_b n^i_a\right)
    \epsilon_b}{\left( \left(n^i_b\right)^2 +
      \left(m^i_b\right)^2\right) I^{(i)}_{ab}}\right) + \nonumber \\ & &
\left( \begin{array}{c} -m_b ^i \\ n_b ^i\end{array}\right) 
\frac{\epsilon_b^i}{\left(n_b^i\right)^2 + \left(m_b^i\right)^2}
  +\left( \begin{array}{c}p_{2i-1}\\p_{2i}\end{array}\right) ,
  \nonumber\\
\left( ac\right)_i &=& \left(\begin{array}{c} n^i_a \label{eq:ipoints}\\
    m^i_a\end{array}\right) \left(\frac{j^{(i)}}{I_{ac}^{(i)}} +\frac{
    \epsilon^i_c}{I^{(i)}_{ca}} -\frac{\left( m^i_a m^i_c + n^i_a n^i_c\right)
    \epsilon_a}{\left( \left(n^i_a\right)^2 +
      \left(m^i_a\right)^2\right) I_{ca}^{(i)}}\right) + \nonumber \\ & &
\left( \begin{array}{c} -m_a ^i \\ n_a ^i\end{array}\right) 
\frac{\epsilon_a^i}{\left(n_a^i\right)^2 + \left(m_a^i\right)^2}
  +\left( \begin{array}{c}q_{2i-1}\\q_{2i}\end{array}\right) ,
  \label{eq:ipinplane}\\
\left(bc\right)_i &=& \left(\begin{array}{c} n^i_c \\
    m^i_c\end{array}\right) \left(\frac{k^{(i)}}{I_{cb}^{(i)}} +\frac{
    \epsilon^i_b}{I^{(i)}_{bc}} -\frac{\left( m^i_b m^i_c + n^i_b n^i_c\right)
    \epsilon_c}{\left( \left(n^i_c\right)^2 +
      \left(m^i_c\right)^2\right) I_{bc}^{(i)}}\right) + \nonumber \\ & &
\left( \begin{array}{c} -m_c ^i \\ n_c ^i\end{array}\right) 
\frac{\epsilon_c^i}{\left(n_c^i\right)^2 + \left(m_c^i\right)^2}
  +\left( \begin{array}{c}t_{2i-1}\\t_{2i}\end{array}\right) ,
  \nonumber
\end{eqnarray}
where $\left( i^{(1)},i^{(2)},i^{(3)}\right)$, $\left( j^{(1)},
  j^{(2)}, j^{(3)}\right)$, $\left( k^{(1)}, k^{(2)}, k^{(3)}\right)$
label inequivalent intersection points between stacks $a$ and $b$,
stacks $a$ and $c$,  stacks $b$ and $c$, respectively. The vectors
$\left( p_1,\ldots , p_6\right)$, $\left( q_1,\ldots ,q_6\right)$ and
$\left( t_1,\ldots , t_6\right)$ belong to the compactification
lattice $\Lambda_{\text{SO(12)}}$. We denote by $z_a^{(i)}$,
$z_b^{(i)}$, $z_c^{(i)}$ ($i \in \left\{ 1,2,3\right\}$)
two-dimensional directional vectors connecting vertices of the
triangle, specifically 
\begin{align}
z_a^{(i)} &= \overrightarrow{\left( ab\right)_i \left( ac\right)_i} ,
\nonumber \\
z_b^{(i)} &= \overrightarrow{\left( bc\right)_i \left( ab\right)_i} ,
\label{eq:edges}\\
z_c^{(i)} &= \overrightarrow{\left( ac\right)_i\left( bc\right)_i}
. \nonumber
\end{align}
These vectors are parallel to the edges of the triangle and have to
add up to zero if the three points (\ref{eq:ipinplane}) are indeed vertices of
a triangle,
\begin{equation} 
z_a^{(i)} + z_b ^{(i)} + z_c ^{(i)} = 0 .
\end{equation}
This condition together with the requirement that the vectors in
(\ref{eq:edges}) are parallel to projections of cycles wrapped by the
stack of D-branes connecting the corresponding intersection points
(projection of 
stack $a$ onto the $i^{\text{th}}$ plane
for $z_a^{(i)}$  and so on) reduce the number of parameters in
(\ref{eq:ipinplane}).  Before imposing these conditions it
proves useful to relabel the intersection points,
\begin{equation}\label{eq:relab}
i^{(i)} \to i^{(i)} I_{ac}^{(i)}/d_a^{(i)},\,\,\,
j^{(i)} \to j^{(i)} I_{cb}^{(i)}/d_c^{(i)},\,\,\,
k^{(i)} \to k^{(i)} I_{ba}^{(i)}/d_b^{(i)} .
\end{equation}
where 
\begin{equation}
d_a^{(i)} = g.c.d.\left( I^{(i)}_{ab} ,  I^{(i)}_{ac}\right) ,\,\,\,
d_b^{(i)} = g.c.d.\left( I^{(i)}_{ba} ,  I^{(i)}_{bc}\right) ,\,\,\, 
d_c^{(i)} = g.c.d.\left( I^{(i)}_{ca} ,  I^{(i)}_{cb}\right) .
\end{equation}

Let us pause to point out a difference to the factorisable $T^6$ where
$d_a^{(i)} = d_b^{(i)} = d_c^{(i)} = d^{(i)}$. This follows from the identity
\begin{equation}
\left( \begin{array}{c} n_b ^i \\ m_b ^i\end{array}\right)
  I^{(i)}_{ac} = I^{(i)}_{ab} \left( \begin{array}{c} n^i _c\\ m
      ^i_c\end{array}\right) + I_{bc}^{(i)}\left( \begin{array}{c} n_a
      ^i \\ m_a ^i\end{array}\right) .
\label{eq:auxid}
\end{equation}
For instance, $n_b ^i$ and $m_b ^i$ are coprime on a factorisable
$T^6$ and hence $I_{ac}^{(i)}$ must be divisible by $g.c.d.\left(
  I^{(i)}_{ab} ,  I_{bc}^{(i)}\right)$. Hence, the greatest common
divisor of all three intersection numbers, $d^{(i)}$, equals the
greatest common divisor of any pair. In the non-factorisable case,
however, it can happen that e.g.\ $g.c.d.\left( n_b^i, m_b^i\right) =
2$ in which case one would conclude $d^{(i)} = d_b^{(i)}/2$. Imagine
for instance that $g.c.d.\left( n_b ^1 ,m_b ^1\right) =2$ whereas
$\left( n_a ^1, m_a ^1\right)$ and $\left( n_c ^1 , m_c^1\right)$ are
pairs of coprime numbers. Then $I_{ab} ^{(1)}$ and $I_{bc}^{(1)}$ are
twice the numbers belonging to the factorisable $T^6$ whereas
$I_{ac}^{(1)}$ does not change. So, in that case $d_b^{(1)} = 2 d^{(1)}$. 

In  terms of the relabelled intersection points the directional vectors in
(\ref{eq:edges}) read
\begin{eqnarray}
z_a^{(i)} &  = &\mathbf{v}_a^i + \left( \begin{array}{c} n_a ^i \\
    m_a^i\end{array}\right)\left( \frac{ 
    j^{(i)} I_{cb}^{(i)}}{d_c ^{(i)} I_{ac}^{(i)}} +\frac{
    \epsilon^i_c}{I^{(i)}_{ca}} -\frac{\left( m^i_a m^i_c + n^i_a n^i_c\right)
    \epsilon_a^i}{\left( \left(n^i_a\right)^2 +
      \left(m^i_a\right)^2\right) I_{ca}^{(i)}}\right)\nonumber\\
& & -\left(  \begin{array}{c} n_b^i \\ m_b^i\end{array}\right) \frac{
    i^{(i)} I_{ac}^{(i)}}{d_a^{(i)} I_{ba}^{(i)}} +
  \left( \begin{array}{c} q_{2i -1} -p_{2i-1}\\ q_{2i}
      -p_{2i}\end{array}\right) , \nonumber \\
z_b^{(i)} &  =& \mathbf{v}_b^i + \left( \begin{array}{c} n_b ^i \\
    m_b^i\end{array}\right)\left( \frac{ 
    i^{(i)} I_{ac}^{(i)}}{d_a ^{(i)} I_{ba}^{(i)}} 
+\frac{
    \epsilon^i_a}{I^{(i)}_{ab}} -\frac{\left( m^i_b m^i_a + n^i_b n^i_a\right)
    \epsilon_b ^i}{\left( \left(n^i_b\right)^2 +
      \left(m^i_b\right)^2\right) I^{(i)}_{ab}}\right) \nonumber\\
& & -\left(  \begin{array}{c} n_c^i \\ m_c^i\end{array}\right) \frac{
    k^{(i)} I_{ba}^{(i)}}{d_b^{(i)} I_{cb}^{(i)}} 
+\left( \begin{array}{c} p_{2i -1} -t_{2i-1}\\ p_{2i}
      -t_{2i}\end{array}\right) ,  \label{eq:zvec}\\
z_c^{(i)} &  = &\mathbf{v}_c^i+\left( \begin{array}{c} n_c ^i \\
    m_c^i\end{array}\right)\left( \frac{ 
    k^{(i)} I_{ba}^{(i)}}{d_b ^{(i)} I_{cb}^{(i)}} 
+\frac{
    \epsilon^i_b}{I^{(i)}_{bc}} -\frac{\left( m^i_b m^i_c + n^i_b n^i_c\right)
    \epsilon_c ^i}{\left( \left(n^i_c\right)^2 +
      \left(m^i_c\right)^2\right) I_{bc}^{(i)}}\right)
\nonumber \\
& &
-\left(  \begin{array}{c} n_a^i \\ m_a^i\end{array}\right) \frac{
    j^{(i)} I_{cb}^{(i)}}{d_c^{(i)} I_{ac}^{(i)}} 
+\left( \begin{array}{c} t_{2i -1} -q_{2i-1}\\ t_{2i}
      -q_{2i}\end{array}\right) ,\nonumber
\end{eqnarray}
with the two dimensional vectors
\begin{eqnarray}
\mathbf{v}_a ^i & = & -\left( \begin{array}{c} n_b^i \\ m_b ^i\end{array}\right)
  \left( \frac{
    \epsilon^i_a}{I^{(i)}_{ab}} -\frac{\left( m^i_b m^i_a + n^i_b n^i_a\right)
    \epsilon_b}{\left( \left(n^i_b\right)^2 +
      \left(m^i_b\right)^2\right) I^{(i)}_{ab}}\right)
+\left( \begin{array}{c}
-m^i _a\\ n^i _a\end{array}\right) \frac{\epsilon^i _a}{\left( n_a
^i\right)^2 +\left( m_a ^i\right)^2} \nonumber \\ & &
-\left( \begin{array}{c}
- m^i _b\\ n^i _b\end{array}\right) \frac{\epsilon_b^i}{\left( n_b
^i\right)^2 +\left( m_b^i\right)^2} ,\nonumber\\
\mathbf{v}_b ^i & = &-\left( \begin{array}{c} n_c ^i \\ m_c
    ^i\end{array}\right) \left( \frac{
    \epsilon^i_b}{I^{(i)}_{bc}} -\frac{\left( m^i_b m^i_c + n^i_b n^i_c\right)
    \epsilon_c}{\left( \left(n^i_c\right)^2 +
      \left(m^i_c\right)^2\right) I_{bc}^{(i)}}\right)
+\left( \begin{array}{c} -m_b^i\\n_b^i\end{array}\right)
\frac{\epsilon_b^i}{\left(n_b ^i\right)^2 +\left( m_b^i\right)^2}
\nonumber \\ & &  -
\left( \begin{array}{c} -m_c ^i\\n_c^i\end{array}\right)
\frac{\epsilon_c ^i}{\left( n_c^i\right)^2 + \left( m_c ^i\right)^2} ,
\\
\mathbf{v}_c^i & = & -\left(\begin{array}{c} n_a ^i\\
    m_a^i\end{array}\right)\left( \frac{
    \epsilon^i_c}{I^{(i)}_{ca}} -\frac{\left( m^i_a m^i_c + n^i_a n^i_c\right)
    \epsilon_a}{\left( \left(n^i_a\right)^2 +
      \left(m^i_a\right)^2\right) I_{ca}^{(i)}}\right)
+\left( \begin{array}{c} -m_c ^i\\n_c^i\end{array}\right)
\frac{\epsilon_c^i}{\left( n_c^i\right)^2 +\left(
    m_c^i\right)^2}\nonumber \\ & & -\left( \begin{array}{c} -m_a^i \\
    n_a ^i\end{array}\right)\
\frac{\epsilon_a ^i}{\left(n_a^i\right)^2 + \left( m_a ^i\right)^2}.\nonumber 
\end{eqnarray}
Conditions on the SO(12) lattice vectors $p$, $q$ and $t$ arise upon
imposing that vectors in (\ref{eq:zvec}) are parallel to two dimensional
projections of the cycles wrapped by the corresponding D-branes. For
instance $z_a^{(i)}$ should be parallel to $\left( n_a^i ,
  m_a^i\right)^{\text{T}}$. Therefore the scalar product of $z_a^{(i)}$ with
$\left( m_a^i,  -n_a^i\right)^{\text{T}}$ has to vanish. This yields
three ($i\in \left\{ 1,2,3\right\}$) linear Diophantine equations
\begin{equation}\label{eq:dio}
n_a^i \left( q_{2i} - p_{2i}\right) -m_a^i\left( q_{2i -1} - p_{2i
    -1}\right) = -\frac{i^{(i)} I_{ac}^{(i)}}{d_a^{(i)}} .
\end{equation}
Each equation is solvable only if the r.h.s. is devisable by
$g.c.d.\left( n_a^{(i)}, m_a^{(i)}\right)$. However, even for
$g.c.d.\left( n_a^{(i)}, m_a^{(i)}\right) =2$ this does not impose
additional conditions since in that case $i^{(i)}$ is even (see e.g.\
case 3 of previous section). There are infinitely many solutions,
\begin{equation}
\left( \begin{array}{c} q_{2i -1} -p_{2i-1} \\
q_{2i}- p_{2i}\end{array}\right) = -\frac{i^{(i)}}{d_a^{(i)}}
\left( \begin{array}{c} n_c^i \\ m_c^i\end{array}\right) + q_a^{(i)}
\left( \begin{array}{c} n_a^i \\ m_a^i\end{array} \right) ,
\label{eq:sec1}
\end{equation}
where $q_a ^{(i)}$ have to be chosen such that $q -
p$ on the l.h.s. is in $\Lambda_{\text{SO(12)}}$. That is, 
the vector on the l.h.s. must have integer components and in addition
the condition
\begin{equation}
\sum_{i=1}^3 \left( -q_a ^{(i)} \left( n_a ^i +  m_a ^i\right) +\frac{
  i^{(i)}\left( n_c ^i + m_c ^i\right)}{d_a^{(i)}}\right) =
0\,\,\,\text{mod}\,\,\, 2 . 
\label{eq:sec2}
\end{equation}
is satisfied.
Also notice that $\mathbf{v}_a ^i$ drops out of the scalar product in
(\ref{eq:dio}) implying that it is parallel to $\left( n_a ^i , m_a
  ^i\right)^\text{T}$ and hence
\begin{equation}
\mathbf{v}_a ^i = \left( \begin{array}{c} n_a ^i\\
    m_a^i\end{array}\right) \frac{\left< \mathbf{v}_a ^i ,
  \left( \begin{array}{c} n_a ^i \\ m_a^i\end{array}\right)\right>}{\left( n_a
      ^i\right)^2 + \left( m_a ^i\right)^2} .
\end{equation}
Using this as well as the identities in (\ref{eq:auxid}) and
\begin{equation}
\frac{ m_a ^i m_c ^i + n_a ^i n_c^i}{I_{ac}^{(i)}} +\frac{ n_a ^i n_b
  ^i + m_a ^i m_b ^i}{I_{ba}^{(i)}} = \frac{\left( \left( n_a
      ^i\right)^2 + \left( m_a ^i\right)^2\right)
  I_{bc}^{(i)}}{I_{ac}^{(i)} I_{ba}^{(i)}}
\end{equation}
one finally finds
\begin{equation}
z_a^{(i)} = \left( \begin{array}{c}n_a^i\\m_a^i\end{array}\right)
I_{bc}^{(i)}\left( 
  \frac{i^{(i)}}{d_a^{(i)}I_{ab}^{(i)}} + \frac{j^{(i)}}{d_c^{(i)}I_{ca}^{(i)}}
      +\frac{I_{bc}^{(i)}\epsilon_a^{i} + I_{ca}^{(i)} \epsilon^i_b
        +I_{ab}^{(i)} \epsilon_c^i}
        {I_{ca}^{(i)} I_{ab}^{(i)} I_{bc}^{(i)}} +
      \frac{q_a^{(i)}}{I_{bc}^{(i)}}\right)   .
\end{equation}
Analogously one finds
\begin{align}
z_b^{(i)} & = \left( \begin{array}{c}n_b^i\\m_b^i\end{array}\right)
I_{ca}^{(i)}\left( 
  \frac{i^{(i)}}{d_a^{(i)}I_{ab}^{(i)}} + \frac{k^{(i)}}{d_b^{(i)}I_{bc}^{(i)}}
      +\frac{I_{bc}^{(i)}\epsilon_a^{i} + I_{ca}^{(i)} \epsilon^i_b
        +I_{ab}^{(i)} \epsilon_c^i}
        {I_{ca}^{(i)} I_{ab}^{(i)} I_{bc}^{(i)}} +
      \frac{q_b^{(i)}}{I_{ca}^{(i)}}\right) , \\  
z_c^{(i)} & = \left( \begin{array}{c}n_c^i\\m_c^i\end{array}\right)
I_{ab}^{(i)}\left( 
  \frac{j^{(i)}}{d_c^{(i)}I_{ca}^{(i)}} + \frac{k^{(i)}}{d_b^{(i)}I_{bc}^{(i)}}
      +\frac{I_{bc}^{(i)}\epsilon_a^{i} + I_{ca}^{(i)} \epsilon^i_b
        +I_{ab}^{(i)} \epsilon_c^i}
        {I_{ca}^{(i)} I_{ab}^{(i)} I_{bc}^{(i)}} +
      \frac{q_c^{(i)}}{I_{ab}^{(i)}}\right) ,
\end{align}
where the $q_b$'s and $q_c$'s satisfy respective conditions,
\begin{align}
-\frac{k^{(i)}}{d_b^{(i)}}\left(\begin{array}{c} n_a ^i \\ m_a
    ^i\end{array}\right)  + q_b^{(i)}\left( \begin{array}{c} n_b ^i \\
    m_b ^i\end{array}\right) & \in {\mathbb Z}^2 ,\label{eq:sec3}\\
\sum_{i=1}^3 \left( -q_b ^{(i)} \left( n_b ^i +  m_b ^i\right) +\frac{
  k^{(i)}\left( n_a ^i + m_a ^i\right)}{d_b^{(i)}}\right) & =
0\,\,\,\text{mod}\,\,\, 2 ,\\
-\frac{j^{(i)}}{d_c ^{(i)}} \left(\begin{array}{c} n_b ^i \\ m_b
    ^i\end{array}\right) + q_c ^{(i)} \left(\begin{array}{c} n_c ^i
  \\ m_c ^i\end{array}\right) & \in {\mathbb Z}^2 ,\\
\sum_{i=1}^3 \left( -q_c ^{(i)} \left( n_c ^i +  m_c ^i\right) +\frac{
  j^{(i)}\left( n_b ^i + m_b ^i\right)}{d_c^{(i)}}\right) & =
0\,\,\,\text{mod}\,\,\, 2.\label{eq:seclast}  
\end{align}
%
Finally, we demand the triangles to close in each plane, i.e.\
\begin{equation}
z_a^{(i)} + z_b ^{(i)} + z_c ^{(i)} =0,\,\,\, \text{for}\,\,\, i \in
\left\{ 1,2,3\right\} .
\end{equation}
For each plane this provides two equations for three variables
$q_a^{(i)}$, $q_b^{(i)}$, $q_c^{(i)}$. Hence there will be three free
parameters $\ell^{(1)}$, $\ell^{(2)}$, $\ell^{(3)}$. 
The solutions are
\begin{align}
q_a^{(i)} & = \frac{k^{(i)}}{d_b^{(i)}} +
\frac{I_{bc}^{(i)}\ell^{(i)}}{d^{(i)}} ,\label{eq:qa}  \\
q_b^{(i)} & = \frac{j^{(i)}}{d_c^{(i)}} + \frac{I_{ca}^{(i)} \ell^{(i)}}{d^{(i)}} ,\\
q_c^{(i)} & = \frac{i^{(i)}}{d_a^{(i)}} + \frac{I_{ab}^{(i)} \ell^{(i)}}{d^{(i)}} ,
\label{eq:qc}
\end{align}
where 
$$ d^{(i)} = g.c.d.\left(d_a^{(i)},d_b ^{(i)},d_c ^{(i)}\right) .$$
Selection rules  resulting from
(\ref{eq:sec1}), (\ref{eq:sec2}) and
(\ref{eq:sec3})--(\ref{eq:seclast}) impose further conditions on these
parameters. We call the set of solutions $\Lambda_3$, i.e.\
\begin{equation}
\left( \ell^{(1)}, \ell^{(2)}, \ell^{(3)}\right) \in \Lambda_3 ,
\end{equation}
$\Lambda_3$ denotes a three dimensional lattice possibly with a lable
dependent off-set.
The worldsheet area of an instanton coupling strings localised at
intersections $i$, $j$, $k$ is (recall each label consists of three
components) 
\begin{align} 
A_{i,j,k}\left( \ell\right) & = \frac{1}{2}\sum_{h=1}^3 \sqrt{ \left|
    z^{(h)}_a\right|^2 
    \left| z_b^{(h)}\right| ^2 -\left( {z_a^{(h)}} ^T
      z_b^{(h)}\right)^2} \nonumber \\ 
& =  \frac{1}{2} \sum_{h=1}^3 \left| I_{ab}^{(h)}
  I_{bc}^{(h)}I_{ca}^{(h)}\right| \left( \frac{i^{(h)}}{d_a^{(h)}I_{ab}^{(h)}} +
  \frac{j^{(h)}}{d_c^{(h)}I_{ca}^{(h)}} + \frac{k^{(h)}}{d_b^{(h)}I_{bc}^{(h)}}
  +\tilde{\epsilon}^{(h)} + \frac{\ell^{(h)}}{d^{(h)}} \right)^2 , 
\end{align}
with
\begin{equation}
\tilde{\epsilon}^{(i)} = \frac{I_{bc}^{(i)}\epsilon_a^{i} + I_{ca}^{(i)} \epsilon^i_b
        +I_{ab}^{(i)} \epsilon_c^i}
        {I_{ca}^{(i)} I_{ab}^{(i)} I_{bc}^{(i)}} .
\end{equation}
Deforming the metric in each of the complex planes  does not affect
the property of $T^6$ to be non-factorisable. This can be easily
included by `covariantising' our expression. Two dimensional scalar
products, as they occur e.g.\ in (\ref{eq:epsdef}) are modified in an
obvious way. We keep $\epsilon^{(i)}$ with the understanding that its
normalisation depends on the metric in the $i^{\text{th}}$
plane. Antisymmetric combinations like $I_{ab}^{(i)}$ transform as
densities and should be multiplied with $\sqrt{g^{(i)}}$, where $g^{(i)}$
is the metric's determinant in the $i^{\text{th}}$ plane. It is
  typically replaced by the K{\"a}hler modulus
\begin{equation}
g^{(i)} = \left(2\pi\right)^2 A^{(i)} .
\end{equation}
Strictly speaking this would be only the K{\"a}hler modulus if we
compactified the two components of $z^{(i)}$ on unit circles. Also in
our case we expect a non-trivial $B$ field with indices in the
$i^{\text{th}}$ plane to amount to a complexified $A^{(i)}$. For the
deformed 2d geometries we obtain
\begin{equation} 
A_{i,j,k}\left(\ell\right) =  \frac{\left( 2\pi\right)^2}{2} \sum_{h=1}^3
A^{(h)}\left| I_{ab}^{(h)} 
  I_{bc}^{(h)}I_{ca}^{(h)}\right| \left( \frac{i^{(h)}}{d_a^{(h)}I_{ab}^{(h)}} +
  \frac{j^{(h)}}{d_c^{(h)}I_{ca}^{(h)}} + \frac{k^{(h)}}{d_b^{(h)}I_{bc}^{(h)}}
  +\tilde{\epsilon}^{(h)} + \frac{\ell^{(h)}}{d^{(h)}} \right)^2 .
\end{equation}
The Yukawa coupling is obtained as a sum over worldsheet instantons
\begin{equation} \label{eq:final}
Y_{ijk} = h_{\text{qu}}\sigma_{abc} \sum_{\ell \in \Lambda_3} \text{exp}\left(
  -\frac{A_{i,j,k}\left(\ell\right)}{2\pi \alpha^\prime}\right) ,
\end{equation}
where 
$$\sigma_{abc} = \text{sign}\left( I_{ab}I_{bc}I_{ca}\right) ,$$
and $h_{\text{qu}}$ is a quantum contribution in accordance with the
corresponding discussion 
in  \cite{Cremades:2003qj}. Their discussion about Wilson lines could be
carried over to the present situation as well. By picking a basis in
$\Lambda_3$ and 
replacing the sum over lattice vectors by a sum over its integer
components our expression for the Yukawa coupling can be brought into
the form of a multi-theta function as anticipated in
\cite{Cremades:2003qj}. 

\section{Examples}

In the present section we will look at two examples. The first example
is designed to focus just on specific
characteristics of branes on non-factorisable six-tori. The second
example will be slightly more complex also featuring subtleties in
cases where some intersection points lose their label in the process
of relabelling.

First, we discuss a very simple setup where particularities due to the
non-factorisable compactification can be demonstrated. We choose the
wrapping numbers according to table \ref{tab:tab1}.
\begin{table}[h]
\begin{center}
\begin{tabular}{c || c| c| c}
plane $(i)$ & 1 & 2 & 3 \\ \hline\hline 
$\left( n_a^i , m_a^i\right)$ & $\left( 2,0\right)$ & $\left( 1,0\right)$
& $\left( 1,0\right)$ \\
\hline
$\left( n_b^i , m_b^i\right)$ & $\left( 0,1\right)$ & $\left( 0,2\right)$
& $\left( 0,1\right) $ \\ 
\hline
$\left( n_c^i , m_c^i\right)$ & $\left( 3,1\right)$ & $\left( 1,1\right)$
& $\left( 1,1\right)$ \\
\hline\hline
$I_{ab}^{(i)}$ & 2 & 2 & 1 \\
\hline
$I_{ab}$ & \multicolumn{3}{|c}{2}\\
\hline 
$I_{ac}^{(i)}$ & 2 & 1 & 1 \\
\hline
$I_{ac}$ & \multicolumn{3}{|c}{1} \\
\hline
$I_{bc}^{(i)}$ & $-3$ & $-2$ & 1 \\
\hline
$I_{bc}$ & \multicolumn{3}{|c}{3}
\end{tabular}
\end{center}
\caption{Cycles and intersection numbers for first
  example. \label{tab:tab1}} 
\end{table}

Inequivalent labels $i$ of  the $ab$-intersection take values in
\begin{equation}
i \in \frac{2{\mathbb Z} \otimes {\mathbb Z}\otimes {\mathbb
    Z}}{\Gamma}
\end{equation}
where $\Gamma$ is a lattice generated by
\begin{equation}
\left( 2,1,0\right) ,\,\,\, \left( 0,2,0\right) ,\,\,\, \left(
  0,1,1\right) .
\end{equation}
There are two equivalence classes which we represent as
\begin{equation}
i \in \left\{\left(  0,0,0\right) , \left( 2,0,0\right)\right\}.
\end{equation}
For the $ac$-intersections there is only one inequivalent label which
we choose as
\begin{equation}
j = \left( 0,0,0\right) .
\end{equation}
Finally, the label of $bc$-intersections $k$  takes values in
\begin{equation}
k \in \frac{{\mathbb Z}\otimes 2{\mathbb Z} \otimes {\mathbb
    Z}}{3{\mathbb Z} \otimes 2 {\mathbb Z} \otimes {\mathbb Z}} .
\end{equation}
Again we can choose representatives of equivalence classes with
vanishing second and third components,
\begin{equation}
k \in \left\{ \left( 0,0,0\right), \left( 1,0,0\right), \left(
    2,0,0\right)\right\} .
\end{equation}
Since all our labels have non-zero entries only in the first component
we replace the three-dimensional vectors by their first components in
the following. Note further, that in our particular example
relabelling according to (\ref{eq:relab}) maps labels to equivalent labels.
The assignment of labels to intersection points is visualised in
figure \ref{fig:ex}.
\begin{figure}[h]
\begin{center}
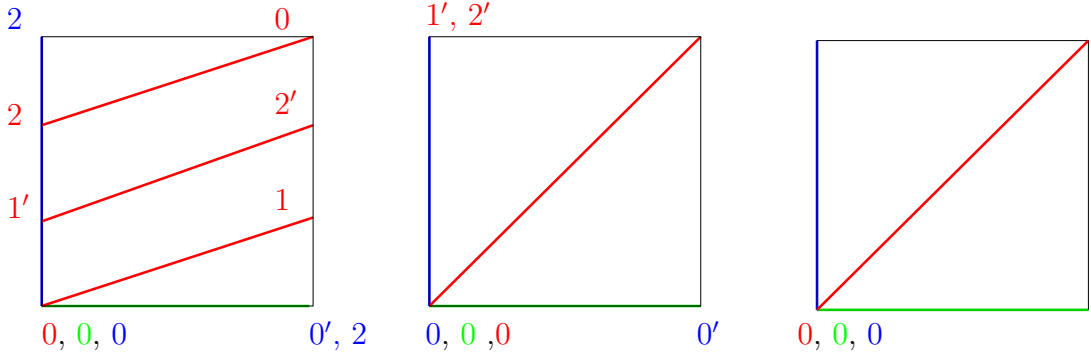
\end{center} 
\caption{A simple example showing subtleties for non-factorisable
  tori. Since inequivalent labels can be chosen to have only
  non-vanishing first components we wrote just the first component. 
Labels are assigned 
  according to our rules. (In the particular example, relabelling does
  not affect them). Values for $i$, $j$, $k$ are 
  depicted in blue, green, red, respectively. Brane stacks $a$, $b$,
  $c$ are drawn in green, blue, read, respectively. Interssections
  with primed labels are related to intersections with the same
  unprimed labels by a lattice shift with non-vanishing components in
  the first and second plane. \label{fig:ex}}
\end{figure}
The largest Yukawa coupling is among fields which are localised at
$i=0$, $j=0$  and $k=0$. Selection rules yield the condition 
\begin{equation}
\left(\begin{array}{c} -6 \ell^{(1)} \\ 0 \\  -2 \ell^{(2)} \\ 0 \\
    \ell^{(3)}\\ 0 \end{array}\right) , 
\left( \begin{array}{c} 0 \\ -2 \ell^{(1)} \\0\\ 2\ell^{(2)} \\ 0\\
    \ell^{(3)}\end{array}\right) ,
\left( \begin{array}{c} 6 \ell^{(1)}\\ 2\ell^{(1)}\\ 2 \ell^{(2)} \\
    2\ell^{(2)} \\\ell^{(3)}\\\ell^{(3)}\end{array}\right) \in
\Lambda_{\text{SO(12)}} .
\end{equation}
They are solved by
\begin{equation}
\ell^{(1)} = \frac{l^{(1)}}{2},\,\,\, \ell^{(2)} = \frac{l^{(2)}}{2}
,\,\,\, \ell^{(3)} = l^{(3)} ,\,\,\, l\equiv\left( \begin{array}{c}
    l^{(1)}\\l^{(2)} \\ l^{(3)} \end{array}\right) \in
\Lambda_{\text{SO(6)}} ,
\end{equation}
where $\Lambda_{\text{SO(6)}}$ consists of three dimensional vectors
  whose integer components sum to an even number. For the Yukawa
  coupling we find
\begin{equation}
Y_{000} = - h_{\text{qu}} \sum_{l \in \Lambda_{\text{SO(6)}}}
\text{exp}\left\{ - \frac{\pi}{\alpha^\prime}\left( 3A^{(1)}
    \left(l^{(1)}\right)^2 + A^{(2)} \left( l^{(2)}\right)
    ^2 + A^{(3)} \left( l^{(3)}\right) ^2\right) \right\} .
\end{equation}
It is also illustrative to compare $Y_{201}$ to $Y_{001}$ which would
be identical on a factorisable $T^6$.  In both cases the selection
rules remain the same as for $Y_{000}$. The couplings can be expressed
as a sum over SO(6) lattice vectors. Instead of writing down the full
instanton sum let us focus on leading contributions, i.e.\ smallest
triangles. We start with $Y_{201}$. In this 
case, the smallest triangle has zero area in the second and third
plane and is depicted in figure \ref{fig:201}. 
\begin{figure}[h]
\begin{center}
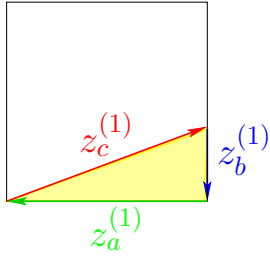
\end{center} 
\caption{The triangle is the worldsheet instanton providing the
  leading contribution to $Y_{201}$. Shown is the projection on the
  first plane. The area on the other two planes is zero. \label{fig:201}}
\end{figure}
This can be easily confirmed by an explicit computation with $ l =0$
\begin{align} 
z_a ^1 & = \left( \begin{array}{c} 2\\ 0\end{array}\right)
  \left( -3\right) 
  \left(\frac{1}{2} -\frac{1}{3}\right) = \left( \begin{array}{c} -1
      \\ 0\end{array}\right) ,\label{eq:za} \\
z_b ^1 & =\left( \begin{array}{c} 0\\ 1\end{array}\right) \left( -2\right)
\left(\frac{1}{2} -\frac{1}{3}\right) = \left( \begin{array}{c} 0\\
    -\frac{1}{3}\end{array}\right) , \\
z_c ^1 & = \left( \begin{array}{c} 3 \\ 1\end{array}\right) 2
\left(\frac{1}{2} -\frac{1}{3}\right) = \left( \begin{array}{c} 1 \\
    \frac{1}{3}\end{array}\right) .\label{eq:zc}
\end{align}
For $Y_{001}$ we find for $l=0$
\begin{align} 
z_a ^1 & = \left( \begin{array}{c} 2\\ 0\end{array}\right)
  \left( -3\right) 
  \left( -\frac{1}{3}\right) = \left( \begin{array}{c} 2
      \\ 0\end{array}\right) , \\
z_b ^1 & =\left( \begin{array}{c} 0\\ 1\end{array}\right) \left( -2\right)
\left( -\frac{1}{3}\right) = \left( \begin{array}{c} 0\\
    \frac{2}{3}\end{array}\right) , \\
z_c ^1 & = \left( \begin{array}{c} 3 \\ 1\end{array}\right) 2
\left( -\frac{1}{3}\right) = \left( \begin{array}{c} -2 \\
    -\frac{2}{3}\end{array}\right) .
\end{align}
The corresponding triangle is drawn in figure \ref{fig:001a}.
\begin{figure}[h]
\begin{center}
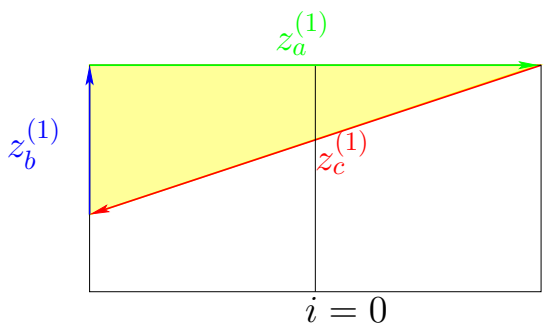
\end{center} 
\caption{The triangle is the worldsheet instanton providing one of the
  leading contribution to $Y_{001}$. Shown is the projection on the
  first plane. The area on the other two planes is zero. \label{fig:001a}}
\end{figure}
In this case there is another competing contribution to the
coupling. It corresponds to the choice 
\begin{equation}
l = \left( \begin{array}{c} -1\\ 1\\ 0\end{array}\right) \to \ell =
\left(\begin{array}{c} -\frac{1}{2}\\ \frac{1}{2} \\ 0\end{array}\right) .
\label{eq:nonzl}
\end{equation}
In the first plane this mimics replacing $i=0$ by $i=2$ and we obtain
the same set of vectors as in (\ref{eq:za}) -- (\ref{eq:zc}). For the
second plane we get
\begin{align}
z_a^{(2)} & = \left(\begin{array}{c} 1 \\ 0\end{array}\right) \left(
    -2\right)\left(  \frac{1}{2}\right) =\left(\begin{array}{c} -1 \\
      0\end{array}\right)  ,\\
z_b^{(2)} & =\left( \begin{array}{c} 0 \\ 2\end{array}\right) \left(
-1\right) \left(\frac{1}{2}\right) = \left(\begin{array}{c} 0 \\
-1\end{array}\right) , \\
z_c ^{(2)} & =\left( \begin{array}{c} 1 \\1 \end{array}\right) 2
\left(\frac{1}{2}\right) 
  = \left( \begin{array}{c} 1 \\ 1\end{array}\right) .
\end{align} 
So, for $\ell$ as in (\ref{eq:nonzl}) the worldsheet instanton action
is given by the sum of triangles in figure \ref{fig:001b}. Depending
on metric moduli this area can be smaller than the $\ell =0$
contribution in figure \ref{fig:001a}.
\begin{figure}[h]
\begin{center}
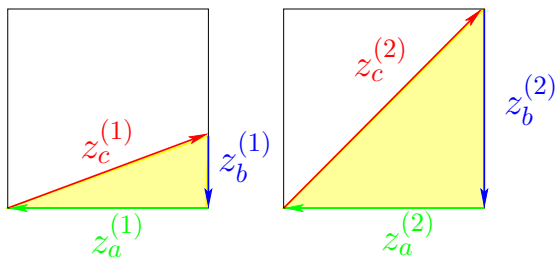
\end{center} 
\caption{The triangle is the worldsheet instanton providing another
  leading contribution to $Y_{001}$.  Shown is the projection on the
  first and second planes. The area on the third plane is
  zero. There is also a worldsheet instanton with the role of second
  and third plane interchanged. \label{fig:001b}}
\end{figure}

As a second example we look at a more generic setup where intersection
points lose their label by relabelling. (A simpler discussion
relevant for factorisable lattices can be found in appendix
\ref{ap:fac}.) Wrapping and intersection
numbers are displayed in table \ref{tab:ex2}.
\begin{table}[h]
\begin{center}
\begin{tabular}{c || c| c| c}
plane $(i)$ & 1 & 2 & 3 \\ \hline\hline 
$\left( n_a^i , m_a^i\right)$ & $\left( 1,-3\right)$ & $\left( 1,1\right)$
& $\left( 1,1\right)$ \\
\hline
$\left( n_b^i , m_b^i\right)$ & $\left( 1,0\right)$ & $\left( 2,0\right)$
& $\left( 1,-2\right) $ \\ 
\hline
$\left( n_c^i , m_c^i\right)$ & $\left( 2,3\right)$ & $\left( 4,6\right)$
& $\left( 1,-1r\right)$ \\
\hline\hline
$I_{ab}^{(i)}$ & 3 & $-2$ & $-3$ \\
\hline
$I_{ab}$ & \multicolumn{3}{|c}{9}\\
\hline 
$I_{ac}^{(i)}$ & 9 & 2 & $-2$ \\
\hline
$I_{ac}$ & \multicolumn{3}{|c}{$-18$} \\
\hline
$I_{bc}^{(i)}$ & 3 & 12 & 1 \\
\hline
$I_{bc}$ & \multicolumn{3}{|c}{18}
\end{tabular}
\end{center}
\caption{Cycles and intersection numbers for second
  example. \label{tab:ex2}} 
\end{table}

Next, we need
to label intersection points. 
First we will follow the prescreption (\ref{eq:ipoints}). 
Let us begin with intersections of
stacks $a$ and $b$.
Intersection points ${\bf p}_{ab}$ are up to $SO(12)$ lattice shifts given
by
\begin{equation}
{\bf
p}_{ab}=\Bigl\{\left(\frac{-3t_1-t_2}{-3},\,0,\,2\frac{t_3-t_4}{2},\,0,\,
\frac{t_5-t_6}{3},\,-2\frac{t_5-t_6}{3}\right)\Big|\vec 
t\in\Lambda_{\text{SO}(12)}\Bigr\}
\end{equation}
We label those points by a triplet  
\begin{equation}
i=\left(-3t_1-t_2,\,t_3-t_4,\,t_5-t_6 \right),
\end{equation}
which is a lattice vector in $\Lambda_{\text{SO}(6)}$.  
Shifting labels as
\begin{equation}
i\rightarrow i+ \left(0,\,2,\,0\right),\quad i\rightarrow i+
\left(3,\,1,\,0\right),\quad i\rightarrow i+ \left(0,\,1,\,3\right).
\end{equation}
leads to equivalent intersection points on $T^6$. Therefore,
inequivalent labels belong to the lattice quotient
\begin{equation}
i\in\frac{\Lambda_{\text{SO}(6)}}{\Gamma_{ab}},
\end{equation}
with
$$
\Gamma_{ab}=\Biggl\{\sum_{i=1}^3n_i\vec e_i\Bigg|n_i\in\mathbb{Z},\,\vec
e_1=\begin{pmatrix}0\\2\\0\end{pmatrix},\,\vec
e_2=\begin{pmatrix}3\\1\\0\end{pmatrix},\,\vec
e_3=\begin{pmatrix}0\\1\\3\end{pmatrix}\Biggr\}.
$$
Relabelling according to (\ref{eq:relab}) corresponds to
\begin{equation}
(i^{(1)},\,i^{(2)},\,i^{(3)})\rightarrow \left(3 i^{(1)},\, i^{(2)},\,
  -2 i^{(3)}\right) , 
\end{equation}
where new labels are again defined up to shifts in $\Gamma_{ab}$. 
If new labels obtained from inequivalent labels become equivalent on
the lattice quotient we do not assign a label to one of the
corresponding intersection points. For instance the label $(1,0,1)$
is mapped to $(3,0,-2)$. Adding the $\Gamma_{ab}$ lattice vector
$\left( -3,0,3\right)$ assigns equivalently the label $\left(
  0,0,1\right)$ which is however already used for the non equivalent
intersection point erstwhile labelled by $\left(
  0,1,1\right)$. Therefore points labelled originally by $\left(
  1,0,1\right)$ lose their label. Analogously one finds the old label
$\left( 2,1,1\right)$ would be also relabelled to $\left(
  0,0,1\right)$. New
labels for $\left( 2,0,2\right)$ and $\left( 1,1,2\right)$ are
equivalent to label $\left( 0,0,2\right)$ reserved for the
relabelled $\left( 0,0,2\right)$. Old labels $\left(
  2,0,0\right)$ and $\left( 1,1,0\right)$ are mapped to $\left(
0,0,0\right)$ which is already taken by the relabelled $\left(
0,0,0\right)$.   
In table \ref{table:tab3} the labels before and after relabelling and the
corresponding coordinates on the torus are listed, a hyphen means
``label lost''.
\begin{table}[h]
\begin{center}
\begin{tabular}{|l|c|l|}\hline\hline
old label $i$ & new label $i^\prime$ & coordinates\\\hline
$(0,\,0,\,0)$&$(0,\,0,\,0)$&$(0,0,0,0,0,0)$\\
$(1,\,0,\,1)$&$-$&$\left(-\frac{1}{3},0,0,0,\frac{1}{3},-\frac{2}{3}\right)$\\ 
$(2,\,0,\,0)$&$-$&$\left(-\frac{2}{3},0,0,0,0,0\right)$\\
$(0,\,0,\,2)$&$(0,\,0,\,2)$&$\left(0,0,0,0,\frac{2}{3},
  -\frac{4}{3}\right)$\\  
$(2,\,0,\,2)$&$-$&$\left(-\frac{2}{3},0,0,0,\frac{2}{3},
  -\frac{4}{3}\right)$\\ 
$(1,\,1,\,0)$&$-$&$\left(-\frac{1}{3},0,1,0,0,0\right)$\\
$(0,\,1,\,1)$&$(0,\,0,\,1)$&$\left(0,0,1,0,\frac{1}{3},-\frac{2}{3}
\right)$\\ 
$(2,\,1,\,1)$&$-$&$\left(-\frac{2}{3},0,1,0,\frac{1}{3},-\frac{2}{3}\right)$\\
$(1,\,1,\,2)$&$-$&$\left(-\frac{1}{3},0,1,0,\frac{1}{3},
-\frac{4}{3}\right)$\\\hline\hline  
\end{tabular}
\end{center}
\caption{Labels and coordinates of intersection points.}
\label{table:tab3}
\end{table}

Next, we investigate intersections of stacks $a$ and $c$. Up to
$SO(12)$ lattice shifts, intersection points are in the set
\begin{equation}
{\bf
p}_{ac}\!=\!\left\{\!\left(\frac{3t_1-2t_2}{9},-3\frac{3t_1-2t_2}{9},
\frac{6t_3-4t_4}{2},\frac{6t_3-4t_4}{2},
\frac{t_5+t_6}{2},\frac{t_5+t_6}{2}\right)\!\left|\vec  
t\in\Lambda_{\text{SO}(12)}\right.\!\!\right\}\! .
\end{equation}
We see that the labels $j=( 3t_1-2t_2,\,6t_3-4t_4  ,\,-t_5-t_6  )$ are
a subset of the 
factorised lattice $\mathbb{Z}\otimes2\mathbb{Z}\otimes\mathbb{Z}$ and
shifting labels by
\begin{equation}
j\rightarrow j+ \left(9,\,0,\,0\right),\quad j\rightarrow j+
\left(0,\,2,\,0\right),\quad 
j\rightarrow j+ \left(0,\,0,\,2\right),
\end{equation}
leaves ${\bf p}_{ac}$ invariant on the torus.
The rule for relabelling components of $j$ is given by
\begin{equation}
j^{(1)} \to -j^{(1)} \,\,\, \text{mod}\,\,\, 9,\,\,\, j^{(2)}\to
-6j^{(2)}\,\,\, \text{mod}\,\,\, 2,\,\,\, j^{(3)} \to -j^{(3)}  \,\,\,
\text{mod}\,\,\,  2.
\end{equation}
In table \ref{table:tab4} one can find the coordinates of labels $j$.
\begin{table}[h]
\begin{center}
\begin{tabular}{|l|l|l|}\hline\hline
old label $j$ & new label $j^\prime$ & coordinates\\\hline
$(0,\,0,\,0)$&$(0,\,0,\,0)$&$\left(0,0,0,0,0,0\right)$\\
$(1,\,0,\,0)$&$(8,\,0,\,0)$&$\left( \frac{1}{9},-\frac{1}{3},0,0,0,0\right)$\\
$(2,\,0,\,0)$&$(7,\,0,\,0)$&$\left(\frac{2}{9},-\frac{2}{3},0,0,0,0\right)$\\
$(3,\,0,\,0)$&$(6,\,0,\,0)$&$\left(\frac{1}{3},-1,0,0,0,0\right)$\\
$(4,\,0,\,0)$&$(5,\,0,\,0)$&$\left(\frac{4}{9},-\frac{4}{3},0,0,0,0\right)$\\
$(5,\,0,\,0)$&$(4,\,0,\,0)$&$\left(\frac{5}{9},-\frac{5}{3},0,0,0,0\right)$\\
$(6,\,0,\,0)$&$(3,\,0,\,0)$&$\left(\frac{2}{3},-2,0,0,0,0\right)$\\
$(7,\,0,\,0)$&$(2,\,0,\,0)$&$\left(\frac{7}{9},-\frac{7}{3},0,0,0,0\right)$\\
$(8,\,0,\,0)$&$(1,\,0,\,0)$&$\left(\frac{8}{9},-\frac{8}{3},0,0,0,0\right)$\\
$(0,\,0,\,1)$&$(0,\,0,\,1)$&$\left(0,0,0,0,\frac{1}{2},\frac{1}{2}\right)$\\
$(1,\,0,\,1)$&$(8,\,0,\,1)$&$\left(\frac{1}{9},-\frac{1}{3},0,0,\frac{1}{2},
  \frac{1}{2}\right)$\\
$(2,\,0,\,1)$&$(7,\,0,\,1)$ &
$\left(\frac{2}{9},-\frac{2}{3},0,0,\frac{1}{2},\frac{1}{2}\right)$ \\
$(3,\,0,\,1)$ & $(6,\,0,\,1)$ &
$\left(\frac{1}{3},-1,0,0,\frac{1}{2},\frac{1}{2}\right)$\\ 
$(4,\,0,\,1)$ & $(5,\,0,\,1)$ &
$\left(\frac{4}{9},-\frac{4}{3},0,0,\frac{1}{2},\frac{1}{2}\right)$ \\
$(5,\,0,\,1)$ & $(4,\,0,\,1)$ &
$\left(\frac{5}{9},-\frac{5}{3},0,0,\frac{1}{2},\frac{1}{2}\right)$ \\
$(6,\,0,\,1)$ & $(3,\,0,\,1)$ &
$\left(\frac{2}{3},-2,0,0,\frac{1}{2},\frac{1}{2}\right)$ \\ 
$(7,\,0,\,1)$ & $(2,\,0,\,1)$ &
$\left(\frac{7}{9},-\frac{7}{3},0,0,\frac{1}{2},\frac{1}{2}\right)$ \\
$(8,\,0,\,1)$ & $(1,\,0,\,1)$ &
$\left(\frac{8}{9},-\frac{8}{3},0,0,\frac{1}{2},\frac{1}{2}\right)$\\
\hline\hline 
\end{tabular}
\end{center}
\caption{Labels and coordinates of intersection points.}
\label{table:tab4}
\end{table}

Finally we need to know which labels to assign to intersections
of stacks $b$ and $c$. Up to lattice shifts intersection points ${\bf
  p}_{bc}$  are given by
\begin{equation}
{\bf p}_{bc}=\left\{\left(2\frac{t_2}{3},\,3\frac{t_2}{3},
  \,4\frac{2t_4}{12},\,6\frac{2t_4}{12},\,2t_5+t_6,\,-2t_5-t_6\right)
\left|\vec t\in\Lambda_{\text{SO}(12)}\right.\right\}. 
\end{equation} 
One can see that labels $k=(t_2 ,\,2t_4
,\,2t_5 +t_6)$ take values on the lattice
$\mathbb{Z}\otimes2\mathbb{Z}\otimes\mathbb{Z}$ and 
intersections points are equivalent on the torus if related by the
following shifts,
\begin{equation}
k\rightarrow k+ \left(0,\,0,\,1\right),\quad k\rightarrow k+
\left(0,\,12,\,0\right),\quad k\rightarrow k+ \left(3,\,6,\,0\right). 
\end{equation}
The set of inequivalent labels is represented by
\begin{equation}
k\in\frac{\mathbb{Z}\otimes2\mathbb{Z}\otimes\mathbb{Z}}{\Gamma_{bc}},
\end{equation}
with
$$
\Gamma_{bc}=\Biggl\{\sum_{i=1}^3n_i\vec
e_i\Bigg|n_i\in\mathbb{Z},\,\vec
e_1=\begin{pmatrix}0\\0\\1\end{pmatrix},\,\vec
e_2=\begin{pmatrix}0\\12\\0\end{pmatrix},\,\vec
e_3=\begin{pmatrix}3\\6\\0\end{pmatrix}\Biggr\}. 
$$
Relabelling according to (\ref{eq:relab}) amounts to redefining
\begin{equation}
\left( k^{(1)},k^{(2)},k^{(3)}\right) \to \left( -k^{(1)},k^{(2)}, 3
  k^{(3)}\right) \,\,\, \text{mod shifts by $\Gamma_{bc}$ lattice
  vectors} . 
\end{equation}
The situation is summarised in table \ref{table:tab5}.
\begin{table}[h]
\begin{center}
\begin{tabular}{|l|l|l|}\hline\hline
old label $k$ & new label $k^\prime$ & coordinates\\\hline
$(0,\,0,\,0)$&$(0,\,0,\,0)$&$\left(0,0,0,0,0,0\right)$\\
$(1,\,0,\,0)$&$(-1,\,0,\,0)$& $\left(-\frac{2}{3}
  ,-1,0,0,0,0\right)$\\ 
$(2,\,0,\,0)$&$(-2,\,0,\,0)$& $\left(-\frac{4}{3},-2,0,0,0,0\right)$\\ 
$(0,\,2,\,0)$&$(0,\,2,\,0)$& $\left(0,0,-\frac{2}{3},-1,0,0\right)$\\
$(1,\,2,\,0)$&$(-1,\,2,\,0)$&
$\left(-\frac{2}{3},-1,-\frac{2}{3},-1,0,0\right)$\\ 
$(2,\,2,\,0)$&$(2,\,2,\,0)$&
$\left(-\frac{4}{3},-2,-\frac{2}{3},-1,0,0\right)$\\ 
$(0,\,4,\,0)$&$(0,\,4,\,0)$& $\left(0,0,-\frac{4}{3},-2,0,0\right)$\\ 
$(1,\,4,\,0)$&$(-1,\,4,\,0)$&
$\left(-\frac{2}{3},-1,-\frac{4}{3},-2,0,0\right)$\\ 
$(2,\,4,\,0)$&$(-2,\,14,\,0)$&
$\left(-\frac{4}{3},-2,-\frac{4}{3},-2,0,0\right)$\\ 
$(0,\,6,\,0)$&$(0,\,6,\,0)$& $\left(0,0,-2,-3,0,0\right)$\\ 
$(1,\,6,\,0)$&$(-1,\,6,\,0)$&
$\left(-\frac{2}{3},-1,-2,-3,0,0\right)$\\ 
$(2,\,6,\,0)$&$(-2,\,6,\,0)$&$\left(-\frac{4}{3},-2,-2,-3,0,0\right)$\\ 
$(0,\,8,\,0)$&$(0,\,8,\,0)$& $\left(0,0,-\frac{8}{3},-4,0,0\right)$\\ 
$(1,\,8,\,0)$&$(-1,\,8,\,0)$&
$\left(-\frac{2}{3},-1,-\frac{8}{3},-4,0,0\right)$\\ 
$(2,\,8,\,0)$&$(-2,\,8,\,0)$&
$\left(-\frac{4}{3},-2,-\frac{8}{3},-4,0,0\right)$\\ 
$(0,\,10,\,0)$&$(0,\,10,\,0)$&
$\left(0,0,-\frac{10}{3},-5,0,0\right)$\\ 
$(1,\,10,\,0)$&$(-1,\,10,\,0)$&
$\left(-\frac{2}{3},-1,-\frac{10}{3},-5,0,0\right)$\\ 
$(2,\,10,\,0)$&$(-2,\,10,\,0)$&
$\left(-\frac{4}{3},-2,-\frac{10}{3},-5,0,0\right)$\\\hline\hline 
\end{tabular}
\end{center}
\caption{Labels and coordinates of intersection points.}
\label{table:tab5}
\end{table}

The data given in the tables \ref{table:tab3}, \ref{table:tab4} and
\ref{table:tab5}  contain every information we need to compute all
trilinear couplings. Our example is designed such that not all
couplings differ from zero. To illustrate that we study selection
rules for couplings to fields belonging to intersection label
$i=\left( 0,0,0\right)$. These were discussed in the previous section
in (\ref{eq:sec1}), (\ref{eq:sec2}), (\ref{eq:sec3}) --
(\ref{eq:seclast}).  For our example they take the form
\begin{equation}\label{condi}
\left(\begin{array}{c}
    q^{(1)}_a\\-3q^{(1)}_a\\q^{(2)}_a\\q^{(2)}_a\\q^{(3)}_a\\q^{(3)}_a  
\end{array}\right)    
,\,\left(\begin{array}{c}
    q^{(1)}_b -\frac{k^{(1)}}{3}\\k^{(1)}\\2q^{(2)}_b
    -\frac{k^{(2)}}{2}\\-\frac{k^{(2)}}{2} \\q^{(3)}_b -
    k^{(3)}\\-2q^{(3)}_b - k^{(3)} \end{array}\right) ,\,
\left(\begin{array}{c}2q^{(1)}_c - \frac{j^{(1)}}{3} \\ 3q^{(1)}_c 
    \\ 4q^{(2)}_c - j^{(2)} \\ 6q^{(2)}_c \\q^{(3)}_c -
    j^{(3)}\\-q^{(3)}_c +2 j^{(3)}\end{array}\right)
\in\Lambda_{\text{SO}(12)}.
\end{equation}
On the other hand conditions for triangles to close fix the $q$'s
to be the form (\ref{eq:qa})--(\ref{eq:qc}) which for our example read
\begin{eqnarray}
q_a^{(1)}=\frac{k^{(1)}}{3}+\ell^{(1)}, &
q_a^{(2)}=\frac{k^{(2)}}{2}+6\ell^{(2)}, &
q_a^{(3)}=k^{(3)}+\ell^{(3)},\nonumber \\ 
q_b^{(1)}=\frac{j^{(1)}}{3}-3\ell^{(1)}, &
q_b^{(2)}=\frac{j^{(2)}}{2}-\ell^{(2)}, &
q_b^{(3)}=j^{(3)}+2\ell^{(3)}, \label{condi2} \\ 
q_c^{(1)}=\ell^{(1)}, & q_c^{(2)}=-\ell^{(2)}, &
q_c^{(3)}=-3\ell^{(3)} . \nonumber 
\end{eqnarray}
Plugging that into (\ref{condi}) yields
conditions on the other labels, $k$ and $j$, as well as on the
$\ell^{(i)}$'s. Imposing the necessary condition on vectors in
(\ref{condi}) to have integer components results in
\begin{align}
\ell^{(1)} & = -\frac{k^{(1)}}{3} + l^{(1)} ,\,\,\, \text{with} \,\,\,
l^{(1)} \in {\mathbb Z} ,\nonumber \\
\frac{j^{(1)}}{3}  & = p + \frac{k^{(1)}}{3},\,\,\, \text{with} \,\,\, p
\in {\mathbb Z} , \nonumber \\
\ell^{(2)} & = \frac{l^{(2)}}{2}\,\,\, \text{with}\,\,\, l^{(2)} \in
{\mathbb Z} , \nonumber \\
\ell^{(3)} & = l^{(3)} \,\,\, \text{with}\,\,\, l^{(3)} \in {\mathbb Z}
. \nonumber 
\end{align}
Note that the second of the above conditions contains a restriction on
possible values for labels $j$ and $k$. Hence, it is really a
selection rule for non vanishing couplings.
Imposing now the full condition (\ref{condi}) yields in addition
$$
l^{(1)} + l^{(2)} + p + j^{(3)} = 0\,\,\, \text{mod} \,\,\, 2 .
$$
So, up to some offset depending on $j^{(3)}
+ p$, the instanton sum will be a sum over the three dimensional lattice
$\Lambda_{\text{SO(4)}} \otimes {\mathbb Z}$, where the
  $\Lambda_{\text{SO(4)}}$ consists of four dimensional vectors with
  integer components whose sum is even. 

As an example we present the coupling between fields lokcalised at
points corresponding to ``new'' labels $i=\left( 0,0,0\right)$,
$j=\left( 8,0,0\right)$ and $k=\left( -1,0,0\right)$,
$$
-h_{\text{qu}}\sum_{l\in(1,\,0,\,0)+\Lambda_{\text{S0}(4)}\otimes
  \mathbb{Z}}\mathrm{exp}\left(-\frac{\pi}{\alpha^\prime}\left[9A^{(1)}
    \left(8+3l^{(1)}\right)^2+12A^{(2)}{l^{(2)}}^2+6A^{(3)}
    {l^{(3)}}^2\right]\right),
$$ 
where $l=\left( l^{(1)}, l^{(2)}, l^{(3)}\right)$ is the worldsheet instanton
  winding number. 
The edges of the area, spread by the instanton, are given by the vectors
\begin{eqnarray}
\begin{pmatrix}z_a^{(1)}\\z_a^{(2)}\\z_a^{(3)}\end{pmatrix}
=  \begin{pmatrix}-\frac{8}{9}   
+l^{(1)}\\\frac{8}{3}-3l^{(1)}\\3l^{(2)}\\
  3l^{(2)}\\l^{(3)}\\l^{(3)}\end{pmatrix},
\quad  \begin{pmatrix}z_b^{(1)}\\z_b^{(2)}\\z_b^{(3)}\end{pmatrix}  
= \begin{pmatrix}2\frac{2}{3}-3l^{(1)}\\0\\-l^{(2)}\\
  0\\2l^{(3)}\\-4l^{(3)}\end{pmatrix},
\quad \begin{pmatrix}z_c^{(1)}\\z_c^{(2)}\\z_c^{(3)}\end{pmatrix}
= \begin{pmatrix}-1\frac{7}{9}+2l^{(1)}\\-\frac{8}{3}+3l^{(1)}
  \\-2l^{(2)}\\-3l^{(2)}\\-3\ell^{(3)}\\3l^{(3)}\end{pmatrix}.  
\end{eqnarray}
The leading contribution to the Yukawa coupling comes either from the
instanton with the winding numbers $l=(1,0,0)$ or $l=(0,1,0)$
depending on the values for $A^{(1)}$ and $A^{(2)}$. In figure
\ref{fig:one} the worldsheet instanton with the winding number
$l=(1,0,0)$ is depicted.  
\begin{figure}[h]
\centering
\resizebox{1\textwidth}{!}{
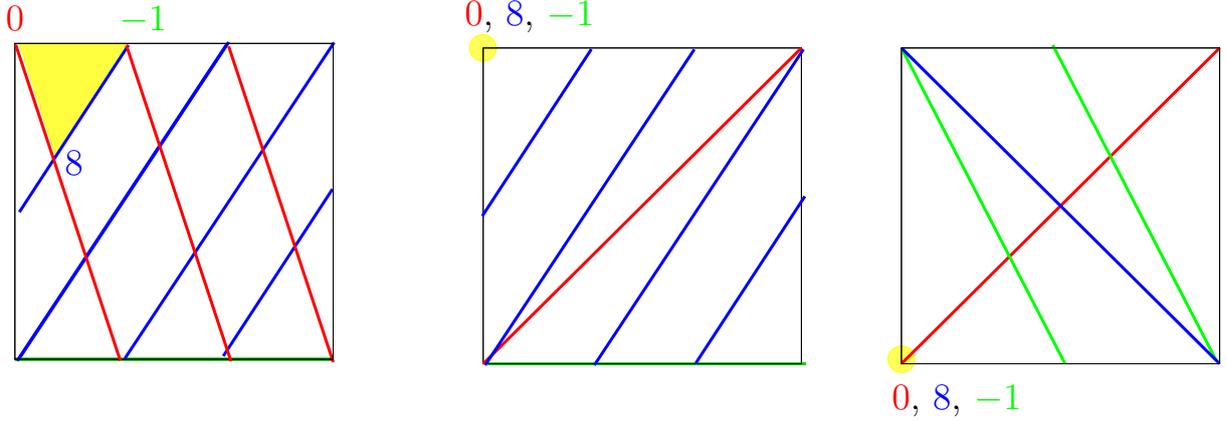}
\caption{The values in red, blue and green are first components of
  labels $i,\,j$ and $k$. The red, green and blue coloured lines
  correspond to the cycles of the branes $a,\,b$ and $c$. The yellow
  area highlighted shows the volume of the instanton with the winding
  number $l=(1,0,0)$ connecting intersections with labels
  $i=(0,0,0),\,j=(8,0,0)$ and $k=(-1,0,0)$ (after
  relabelling). \label{fig:one}} 
\end{figure}

As we have seen, intersections sometimes lose their label in the
process of relabelling. Still, we can compute Yukawa couplings
involving fields localised at such intersection. The general strategy
is as follows. First, consider the intersection which loses its label
in the process of relabelling. We shift the old label by a fraction of
an equivalence shift. The fraction is determined by the greatest common
divisor of the three intersection numbers. The shifted label should be
mapped to an existing new label by relabelling. The other labels need
in general also to be shifted by the same fraction of corresponding
equivalence shifts. Which particular equivalence shift should be
taken is determined by the requirements that the shifted label exists,
and its relabelled version exists. Further, associated coordinate
shifts should coincide on ${\mathbb R}^6/\Lambda_{\text{SO(12)}}$
for all three intersections.  
If this way selection rules cannot be satisfied the corresponding
Yukawa coupling is zero.  We illustrate the general prescription at an
example. 

Consider three intersections with original labels $i=\left(
  2,0,0\right)$, $j =\left( 4,0,0\right)$ and $k=\left(
  2,6,0\right)$. We indicate the positions of the corresponding
intersections in figure \ref{fig:lost}.
\begin{figure}[h!]
\centering
\resizebox{1\textwidth}{!}{
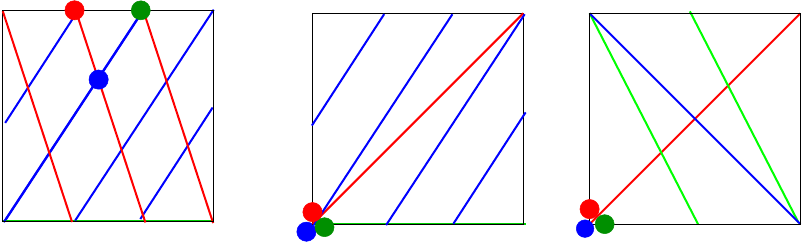}
\caption{The intersection marked by the red disc loses its label in
  the process of relabelling. The Yukawa coupling to fields localised
  at intersections marked by blue and green discs can still be
  computed and is identical to the previously considered
  coupling. \label{fig:lost}}  
\end{figure}
The intersection point of stacks $a$ and $b$ loses its label in the
process of relabelling. 
Using our general prescription we change labels by third fractions of
equivalence shifts, explicitly given by
\begin{equation}
i \to i - \frac{1}{3}\left( 6,0,0\right) ,\,\,\, j \to
j-\frac{1}{3}\left( 9,0,0\right) ,\,\,\, k \to k -\frac{1}{3}\left(
  3,18,0\right) .
\end{equation}
The shifted labels are identical to the original labels of our
previously computed coupling. As expected, we obtain the same coupling, now.

\section{Conclusions}

The major subject of the current paper is the computation of Yukawa
couplings in intersecting brane models on non-factorisable tori. The
result can be found in Eq.\ (\ref{eq:final}). For the factorisable
torus a similar expression is given in  \cite{Cremades:2003qj}. 
The most significant difference is that the latter can be written as a
product of three theta functions whereas (\ref{eq:final}) contains a
multi-theta function. 
For our calculation we represent D-branes in a way closely related to
branes on factorisable $T^6$. We carefully address
the question of how to label intersection points. It turns out that a
label in general takes values on a three dimensional quotient
lattice. Our original prescription of assigning labels to
intersections depends only on the two branes involved but is
asymmetric under permuting them. For the computation of Yukawa
couplings it is convenient to re-assign new labels depending now even
on the third brane. After such relabelling the computation of Yukawa
couplings leads to Diophantine equations of a particular form. They are
equivalent to equations arrising in the problem of finding for a given
brane a partner brane such that their intersection number takes a given
value. Moreover, branes and intersection numbers correspond to ones
already given by the setup. Therefore the general solution can be
given in terms of wrapping numbers of the model and quantities
reflecting the fact that parallel branes have zero intersection
number. The number of these additional quantities is further reduced
by more concistency conditions and one is left with three dimensional
lattice vectors. These can be viewed as labelling worldsheet
instantons, i.e.\ they can be understood as generalised wrapping
numbers for open string worldsheet instantons.  If these wrapping
numbers take values on a factorisable lattice the Yukawa coupling will
be expressed in terms of a product of three theta
functions. Generically this is not possible.  

We discuss subtleties for intersection numbers having
non-trivial common divisors. We find that our way of relabelling
intersections is not always bijective in such cases, i.e.\ some
intersections lose their label. On the other hand there is a
degeneracy in Yukawa couplings, i.e.\ different cubic interactions
have identical coupling constants. This fact allows to relate
couplings for fields with no label to others which can be computed.  

We hope our work will contribute to efforts in intersecting brane
model building. It should help to extend existing strategies to the
study of non-factorisable tori. Recent attempts in that direction have
been reported in \cite{Bailin:2013sya}. There,  ${\mathbb Z}_{12}$
orientifolds with chiral spectra are investigated. It will be
interesting to find out whether techniques developed in our paper can
also be used in that context. Certainly, our methods are easily adopted
to all orientifolds whose point group can be also realised as an
automorphism of a factorisable six-torus.

\section*{Acknowledgements}  
This work was supported by the SFB-Transregio TR33 ``The Dark
Universe'' (Deutsche Forschungsgemeinschaft) and by ``Bonn-Cologne
Graduate School for Physics and Astronomy'' (BCGS).

\appendix
\section{{$\mathbf T^2$} with Non-Coprime Intersection 
Numbers\label{ap:fac}} 

To deal with the case of non-coprime intersection numbers the authors
of \cite{Cremades:2003qj} used a well motivated general ansatz
together with a case by case study to compute Yukawa couplings. 
The second step is best performed by drawing a picture and fitting it
with the general ansatz. 
Here, we propose an alternative treatment allowing for a more
formalised deduction of the same results. That will involve
relabelling intersection points in such a way that some 
inequivalent labels are lost. Yukawa couplings containing
corresponding fields are equal to other Yukawa couplings. These
contain fields belonging to intersection points retained after
relabelling. Since we are considering just one $T^2$  factor we drop
the index labelling the three complex planes. Further we consider
the unit square lattice and the case that all branes pass through the
origin. The analog of equations (\ref{eq:ipoints}) reads
\begin{align}
\left( ab\right) & = \left( \begin{array}{c} n_b \\
    m_b\end{array}\right) \frac{i}{I_{ba}} + \left(\begin{array}{c}
    p_1 \\ p_2\end{array}\right) , \nonumber \\
\left( ac\right) & = \left( \begin{array}{c} n_a \\
    m_a\end{array}\right) \frac{j}{I_{ac}} +\left( \begin{array}{c}
    q_1 \\ q_2\end{array}\right) , \label{eq:bre}\\
\left( bc\right) & = \left( \begin{array}{c} n_c \\
    m_c\end{array}\right) \frac{k}{I_{cb}} + \left( \begin{array}{c}
    t_1\\t_2\end{array}\right) , 
  \nonumber
\end{align}
where now the $p_\alpha$, $ q_\alpha$, $t_\alpha$ are just integers
with no further constraints. The labels $i$, $j$, $k$ are defined up
to shifts by integer multiples of $I_{ba}$, $I_{ac}$ and $I_{bc}$,
respectively. Then, as explained in section
\ref{sec:yuk} the greatest common divisor for any pair of
intersection numbers is equal to
\begin{equation}
d = g.c.d.\left( I_{ab}, I_{ac}, I_{bc}\right) .
\end{equation}
The relabelling in (\ref{eq:relab}) simplifies to
\begin{equation}
i \to i I_{ac}/d,\,\,\, j \to  j I_{cb}/d ,\,\,\, k \to k I_{ba}/d
\label{eq:re}
\end{equation}
yielding
\begin{align}
\left( ab\right) & = \left( \begin{array}{c} n_b \\
    m_b\end{array}\right) \frac{i I_{ac}}{d I_{ba}} + \left(\begin{array}{c}
    p_1 \\ p_2\end{array}\right) , \nonumber \\
\left( ac\right) & = \left( \begin{array}{c} n_a \\
    m_a\end{array}\right) \frac{j I_{cb}}{d I_{ac}} +\left( \begin{array}{c}
    q_1 \\ q_2\end{array}\right) , \\
\left( bc\right) & = \left( \begin{array}{c} n_c \\
    m_c\end{array}\right)\frac{ k I_{ba}}{dI_{cb}} +
\left( \begin{array}{c} t_1\\t_2\end{array}\right) .
  \nonumber
\end{align}
In general it will happen that this expression misses some
intersection points if e.g.\ $I_{ac}/d$ is devisable by $I_{ab}$. 
We proceed by computing directional vectors connecting intersection
points
\begin{align}
z_a & = \overrightarrow{  \left( ab\right) \left( ac\right)} =
\left( \begin{array}{c} n_a \\ m_a \end{array}\right) \frac{j
  I_{cb}}{d I_{ac}} - \left( \begin{array}{c} n_b \\
    m_b\end{array}\right) \frac{i I_{ac}}{d I_{ba}} + \left( \begin{array}{c}
    q_1 -p_1\\ q_2 - p_2\end{array}\right) , \nonumber \\
z_b & = \overrightarrow{\left( bc\right) \left( ab\right)} =
\left( \begin{array}{c} n_b \\ 
m_b\end{array}\right) \frac{i I_{ac}}{d I_{ba}} -
\left( \begin{array}{c} n_c \\
    m_c\end{array}\right)\frac{ k I_{ba}}{dI_{cb}} + \left(\begin{array}{c}
    p_1 -t_1\\ p_2 -t_2\end{array}\right) , \\
z_c & = \overrightarrow{\left( ac\right) \left( bc\right) } =
 \left( \begin{array}{c} n_c \\
    m_c\end{array}\right)\frac{ k I_{ba}}{dI_{cb}} -
\left( \begin{array}{c} n_a \\
    m_a\end{array}\right) \frac{j I_{cb}}{d I_{ac}} +
\left( \begin{array}{c} t_1 -q_1 \\t_2 - q_2\end{array}\right) .
\nonumber
\end{align}
Imposing $z_x$ to be parallel to $\left( n_x , m_x\right)^{\text{T}}$
for $x \in \left\{ a,b,c\right\}$ gives linear Diophantine equations
with solutions
\begin{align}
\left( \begin{array}{c} q_1 - p_1 \\ q_2 - p_2\end{array}\right) & =
\left( \begin{array}{c} n_a \\ m_a\end{array}\right) q_a
-\left( \begin{array}{c} n_c\\ m_c\end{array}\right) \frac{i}{d} ,
\nonumber \\
\left( \begin{array}{c} p_1 - t_1\\ p_2 - t_2\end{array}\right) & =
\left( \begin{array}{c} n_b\\m_b\end{array}\right) q_b -
\left( \begin{array}{c} n_a \\ m_a\end{array}\right) \frac{k}{d}
,\label{eq:sel} \\ 
\left( \begin{array}{c} t_1 - q_1 \\ t_2 - q_2\end{array}\right) & =
\left( \begin{array}{c} n_c \\ m_c\end{array}\right) q_c
-\left( \begin{array}{c} n_b \\ m_b\end{array}\right) \frac{j}{d} ,
\nonumber
\end{align}
where $q_a$, $q_b$ and $q_c$ are parameters which are related via the
condition for the triangle to close
\begin{equation}
z_a + z_b + z_c = 0.
\end{equation}
These provide two equations for the three parameters $q_x$. We
parameterise the solution by $l_0 + l$ where $l_0$ denotes a
fractional part and $l$ an integer. 
For the $q_x$ one finds
\begin{equation}
q_a = \frac{k}{d} +\frac{\left( l_0 +l\right) I_{bc}}{d}\,\,\, ,
q_b = \frac{j}{d} + \frac{ \left(l_0 +l\right) I_{ca}}{d},\,\,\, q_c =
\frac{i}{d} +\frac{\left( l_0 + l\right) I_{ab}}{d}
\end{equation}
Then the $z_x$ can be written as
\begin{equation}
z_a =\left( \begin{array}{c}n_a\\ m_a\end{array}\right) \frac{I_{bc}}{d}\left(
  x_0 + l\right) ,  \,\,\,
 z_b =\left( \begin{array}{c}n_b\\ m_b\end{array}\right) \frac{I_{ca}}{d}\left(
  x_0 + l\right)  ,\,\,\,
z_c =\left( \begin{array}{c}n_c\\ m_c\end{array}\right) \frac{I_{ab}}{d}\left(
  x_0 + l\right) ,
\label{eq:zets}
\end{equation}
with
\begin{equation}
x_0 = \frac{i}{I_{ab}} + \frac{j}{I_{ca}} + \frac{k}{I_{bc}} + l_0 ,
\end{equation}
which is defined up to integer shifts.
One still has to take into account that the vectors on the left hand
sides of (\ref{eq:sel}) have integer entries. This provides selection
rules on possible Yukawa couplings.  The integer parameter
$l$ contributes only integer numbers to the right hand sides and just
drops off the selection rules. For $l_0$ running through $\frac{1}{d},
\frac{2}{d}, {\dots}, \frac{d-1}{d}$ one will get different selection
rules on the intersection labels. Thus $l_0$, indeed, represents a
contribution to 
$x_0$ of the form $s\left( i,j,k\right)/d$  as postulated in
\cite{Cremades:2003qj}. 

To illustrate our general discussion we revisit an example discussed
in \cite{Cremades:2003qj}. The wrapping numbers are
\begin{equation}
\left( n_a, m_a\right) = \left( 1,0\right) ,\,\,\, \left( n_b,
  m_b\right) = \left( 1,2\right),\,\,\, \left( n_c, m_c\right) =
\left( 1, -4\right) .
\end{equation}
The setup is depicted in figure \ref{fig:a1}.
\begin{figure}
\begin{center}
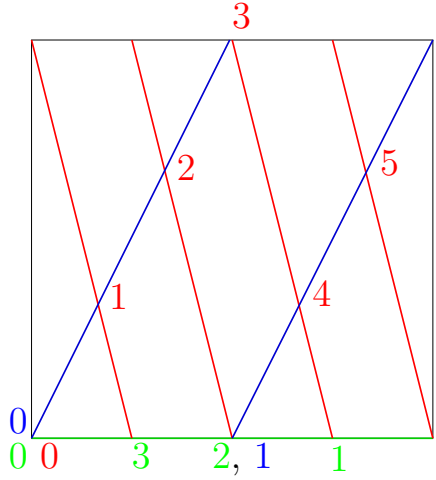
\end{center} 
\caption{Setup from \cite{Cremades:2003qj}. Labels are assigned
  according to our rules before relabelling. Values for $i$, $j$, $k$ are
  depicted in blue, green, red, respectively. Brane stacks $a$, $b$,
  $c$ are drawn in green, blue, read, respectively. \label{fig:a1}}
\end{figure}
Here labels are allocated to intersection points according to
(\ref{eq:bre}). Relabelling as in (\ref{eq:re}) amounts to the
replacements
$$
i: \left( 0,1\right) \to (0, \text{no label}), \,\,\, j: \left(
  0,1,2,3\right) \to 
\left( 0,3,2,1\right) ,\,\,\, k: \left( 0,1,2,3,4,5\right) \to \left(
  0,5,4,3,2,1\right) , 
$$
where we notice that the intersection point previously labelled
by $i=1$ does not have a label any more. However, for any triangle
containing a vertex labelled by $i=1$ one can find a congruent
triangle with vertex labelled by $i=0$. The selection rules (taken
from (\ref{eq:sel})) imply that $k+j$ has to be even, always. For
$l_0 =0$, $k$ and $j$ must be even whereas for $l_0 = \frac{1}{2}$
they must be odd. The situation can be summarised in the selection
rule
\begin{equation}
k+j = 0\,\,\, \text{mod}\,\,\, 2 ,
\end{equation}
together with (recall $x_0$ is defined modulo one)
\begin{equation}
x_0 = \frac{j}{4} - \frac{k}{6} + \frac{j}{2} =\frac{3j}{4} -\frac{k}{6}.
\label{eq:x0}
\end{equation}
This agrees with the result reported in
\cite{Cremades:2003qj} as long as $i=0$. Now suppose, we want to obtain the
Yukawa coupling for the triplet $\left( i,j,k\right) = (1,1,4)$
in figure \ref{fig:a1}. Since the intersection point labelled by
$i=1$ loses its label in the process of relabelling we first perform a
shift 
\begin{equation}
\left( i, j, k\right) \to
\left( i, j, k\right) -
\frac{1}{d}\left( I_{ba}, I_{ac} ,
    I_{cb}\right) ,
\end{equation}
i.e.\ by $\left( 1 , 2, -3\right)$. This tells us that the Yukawa
coupling of fields localised at $( 0, 3, 1)$ is the same. Relabelling    
maps this finally to $\left( i,j,k\right)=\left( 0,1,5\right)$. With
(\ref{eq:x0}) and (\ref{eq:zets}) we obtain for $l=0$  
$$ x_0 = -\frac{1}{12},\,\,\, z_a = \left( \begin{array}{c}  \frac{1}{4} \\
    0\end{array} \right) ,\,\,\, z_b = \left(\begin{array}{c}
    -\frac{1}{6} \\ -\frac{1}{3}\end{array}\right) ,\,\,\, z_c =
  \left( \begin{array}{r} -\frac{1}{12}\\ \frac{1}{3}\end{array}
  \right) , $$ 
in accordance with figure \ref{fig:a1}.

\end{document}